\documentclass[a4paper,11pt]{article}
\usepackage{jheppub} 
\usepackage{xcolor}
\definecolor{red}{rgb}{0,0,0}
\usepackage{slashed}
\usepackage{subcaption}

\makeatletter
\def\@fpheader{\ }
\makeatother

\title{\boldmath \boldmath The near-threshold cross section of $e^+e^- \to \Omega^-\bar{\Omega}^+$: Heavy-flavor rescattering and physics-informed deep learning}

\author{Sara Rahmani}
\affiliation{Departamento de Física Aplicada, \textcolor{red}{Cinvestav}, Carretera Antigua a Progreso Km. 6, Apdo. Postal 73 “Cordemex”, 97310 Mérida, Yucatán, \textcolor{red}{México}}

\emailAdd{sara.rahmani@cinvestav.mx}

\abstract{Recent precision measurements of the $e^+e^- \to \Omega^-\bar{\Omega}^+$ cross section by the BESIII collaboration provide a valuable opportunity to probe complex hadronic rescattering mechanisms. In this work, we investigate a potential structure near the  $D_s\bar{D}_s$ threshold using a coupled-channel framework incorporating $\Omega\bar{\Omega}$, $\Xi\bar{\Xi}$, and $D_s\bar{D}_s$ interactions. The driving potentials are derived from effective Lagrangians respecting heavy quark spin symmetry, chiral symmetry, and hidden local symmetry, and the scattering amplitude is unitarized via the on-shell factorization of the Bethe-Salpeter equation. 
To go beyond local fits and map theoretical uncertainties, we use a two-step machine-learning framework. First, Simulation-Based Inference with a Mixture Density Network maps the global Bayesian posterior of the effective couplings. Second, \textcolor{red}{to identify the poles,} we employ a Cauchy-Riemann Physics-Informed Neural Network (PINN). The network enforces mathematical analyticity \textcolor{red}{by
construction},  continuing the real-axis \textcolor{red}{determinant} into the complex energy plane. \textcolor{red}{At the best fit, no pole is found near the $D_s\bar D_s$ threshold, while a single physical pole is identified  at $M=3.1868$ GeV with $\Gamma=1.57$ MeV, sitting
158 MeV below the $\Omega\bar\Omega$ threshold.} The corresponding
S-matrix residues show an overwhelming coupling to the \textcolor{red}{$\Omega\bar\Omega$} channel,
\textcolor{red}{favoring a} dynamically generated \textcolor{red}{$\Omega\bar\Omega$} bound state\textcolor{red}{, predicted by the model below the measured energy range}.}

\begin{document}
\maketitle
\flushbottom

\section{Introduction}
\label{sec:intro}
In recent years, high-precision measurements of hadron production in electron-positron collisions near kinematic thresholds have revealed a rich spectrum of unexpected enhancements and exotic structures. In particular, the BESIII collaboration has measured the Born cross section and effective form factors for hyperon-antihyperon pair production processes, such as $e^+e^- \to \Omega^-\bar{\Omega}^+$~\cite{BESIII:2022kzc, BESIII:2025fph, BESIII:2026qyk}. Determining whether such structures originate from genuine charmonium-like resonances, kinematic threshold effects, or dynamically generated hadronic molecules is a central topic in modern hadron spectroscopy~\cite{Zhang:2025qmo,Haidenbauer:2020wyp,Dai:2017fwx,Jia:2024ybo,Zhang:2023wmd,Salnikov:2023qnn}.
Milstein and Salnikov investigated the near-threshold energy dependence of the $e^+e^- \to \Lambda_c  \bar{\Lambda}_c$
cross section by incorporating final-state interactions, \textcolor{red}{showing that these interactions account for the
observed energy dependence~\cite{Milstein:2022bfg} and predicting a narrow $\Lambda_c\bar\Lambda_c$ bound state
approximately 38~MeV below threshold \cite{Salnikov:2023qnn}}. A recent coupled-channel analysis by Jia \textit{et al.} demonstrated that strong final-state interactions in the
$e^+e^- \to N\bar{N}$
 process successfully explain near-threshold line shapes by dynamically generating $N\bar{N}$ quasibound states \cite{Jia:2024ybo}.
Indeed, a recent phenomenological analysis by Zhang and Wang searched for direct charmonium(-like) resonance contributions to the $e^+e^- \to \Omega^-\bar{\Omega}^+$ cross section but found no significant signals, \textcolor{red}{and the resulting upper limits on the branching
fractions remain at least an order of magnitude above the perturbative QCD
scaling \cite{Zhang:2025qmo}. These findings leave the origin of the observed line shape open, motivating further investigation of} a heavy-flavor rescattering mechanism. 

Unitarized coupled-channel frameworks based on effective chiral Lagrangians have proven to be powerful tools for understanding near-threshold phenomena~\cite{Hyodo:2011ur, Doring:2025sgb}. In these approaches, strong interactions between hadrons can dynamically generate bound states or resonances as poles in the scattering matrix, without requiring pre-existing quark-model states. This picture has been successfully applied to describe a wide variety of strange and heavy-flavor hadrons~\cite{Qi:2023gwb, Shen:2022zvd, Kim:2025ado, Qi:2021iyv, Zhang:2024fxy, Moir:2016srx}. In $e^+e^-$
 annihilation, prompt hyperon pair production is described by timelike electromagnetic form factors~\cite{Lih:2026xgi}. Near threshold, this direct process is complemented by strong final-state rescattering through intermediate channels such as $|\Xi\bar{\Xi}\rangle$
 and heavy-meson pairs like $|D_s^-\bar{D}_s^+\rangle$.
\textcolor{red}{The production of open-charm meson pairs, including $D_s^-D_s^+$ channel, has been measured
by BABAR, Belle, CLEO-c and BESIII. The measured cross sections provide important input for
coupled-channel analyses \cite{Wang:2025dur}.}

While unitarized models offer a firm theoretical foundation, extracting physical couplings and mapping theoretical uncertainties from experimental data remains challenging.
Complementing conventional $\chi^2$ minimization techniques \cite{Pilloni:2016obd}, machine learning methods have recently emerged as valuable tools in hadron spectroscopy for analyzing complex amplitude structures, extracting $S$-matrix poles, and quantifying parameter uncertainties~\cite{Frohnert:2025usi,Sombillo:2021rxv,Ng:2021ibr,Aarts:2025gyp,Liu:2022uex}. In particular, when quantum interference between direct production and rescattering loops creates parameter trade-offs, Simulation-Based Inference (SBI) \cite{Cranmer:2019eaq} offers an effective way to map the full Bayesian posterior probability landscape\textcolor{red}{, with the nuisance parameters marginalized. For a relatively small parameter space, a standard sampler can yield the same posterior, so we benchmark SBI against nested
sampling using the same likelihood and priors. The advantage of SBI is that, once trained, it evaluates
the posterior for any new data set at negligible computational cost.}

Once the parameter landscape is mapped, uncovering the physical origin of these line-shape features requires \textcolor{red}{probing for} the complex $S$-matrix pole position ($E_{\mathrm{pole}} = M - i\Gamma/2$). 
\textcolor{red}{once the couplings
are fixed, the amplitude is fully specified, allowing us to determine the poles directly by finding the zeros of $\det[1-VG]$ on the relevant Riemann
sheets. As an independent approach that relies only on amplitude values along the real axis,} we construct a Cauchy-Riemann Physics-Informed Neural Network (PINN) as a global holomorphic surrogate that performs analytic continuation off the real axis \textcolor{red}{, and we validate this approach against the
direct pole search, employing an ensemble of networks to quantify the uncertainty of the analytic continuation}.

In this paper, we investigate the $e^+e^- \to \Omega^-\bar{\Omega}^+$ cross section using a $3 \times 3$ unitarized coupled-channel model ($|\Omega^-\bar{\Omega}^+\rangle$, $|\Xi\bar{\Xi}\rangle$, $|D_s^-\bar{D}_s^+\rangle$). The remainder of this manuscript is organized as follows. In Section~\ref{sec: formalism}, we construct the tree-level transition potentials and unitarize the scattering amplitude via the Bethe-Salpeter equation. Section~\ref{sec:results} presents the phenomenological fit to BESIII data, the global Bayesian uncertainty mapping via SBI \textcolor{red}{together with the corresponding nested-sampling benchmark}, and the pole extraction using \textcolor{red}{the direct pole search and} our Cauchy-Riemann PINN. Finally, Section~\ref{sec: summary} provides a summary of our conclusions and an outlook for future applications.

\section{Formalism}
\label{sec: formalism}
We investigate the $e^+e^- \to \Omega^-\bar{\Omega}^+$ production cross section in the near-threshold region, explicitly incorporating the dynamical effects of heavy-flavor coupled-channel rescattering. The initial state is produced via $e^+e^-$ annihilation into a virtual photon, imposing conservation of the quantum numbers $J^{PC} = 1^{--}$ on the final state system.
The Born cross section for $e^+e^- \to \Omega^-\bar{\Omega}^+$ can be expressed in \textcolor{red}{in
terms of the phenomenological amplitude $F(s)$ as:}
\begin{equation}
    \sigma(e^+e^- \to \Omega^-\bar{\Omega}^+) = \frac{4\pi\alpha^2}{3s} \frac{|\vec{p}_\Omega|}{\sqrt{s}} |F(s)|^2, \label{eq:sigma}
\end{equation}
where $\alpha$ is the fine-structure constant, $s$ is the center-of-mass energy squared, and $|\vec{p}_\Omega|$ is the center-of-mass three-momentum of the outgoing $\Omega^-$ baryon.

\begin{figure}[htbp]
    \centering
    
    \begin{subfigure}[b]{0.48\textwidth}
        \centering
        \includegraphics[width=\textwidth]{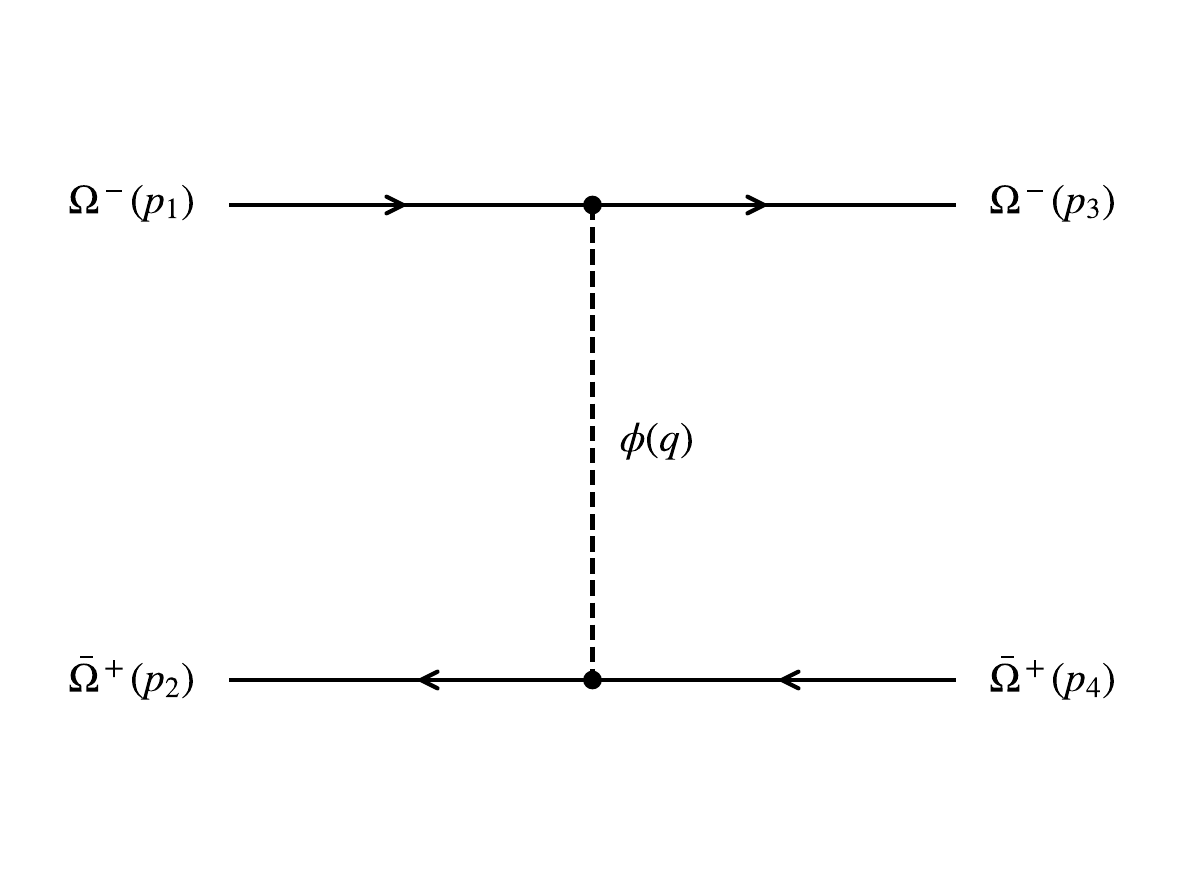}
        \caption{$V_{11}$: $\Omega^-\bar{\Omega}^+ \to \Omega^-\bar{\Omega}^+$}
        \label{fig:V11}
    \end{subfigure}
    \hfill
    \begin{subfigure}[b]{0.48\textwidth}
        \centering
        \includegraphics[width=\textwidth]{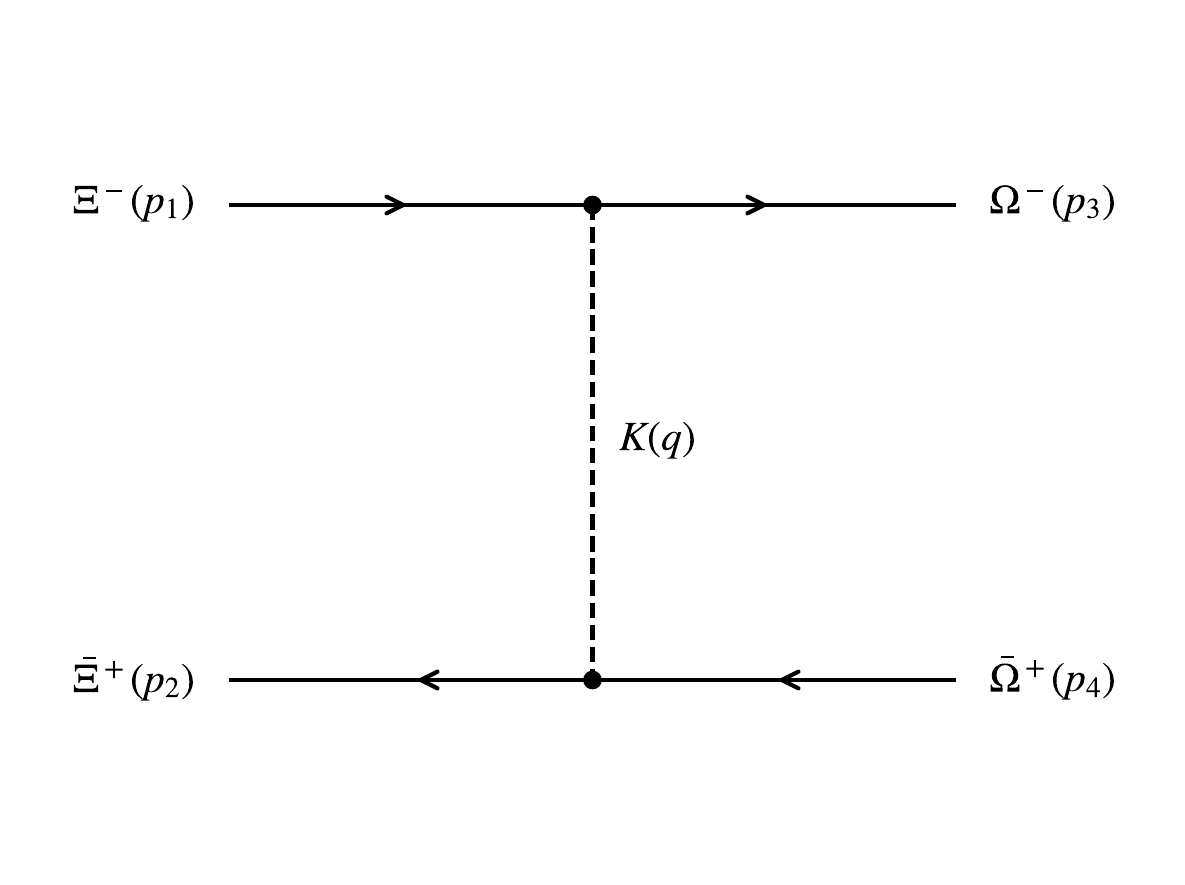}
        \caption{$V_{12}$: $\Xi^-\bar{\Xi}^+ \to \Omega^-\bar{\Omega}^+$}
        \label{fig:V12}
    \end{subfigure}
    
    \vspace{0.6cm} 
    
    \begin{subfigure}[b]{0.48\textwidth}
        \centering
        \includegraphics[width=\textwidth]{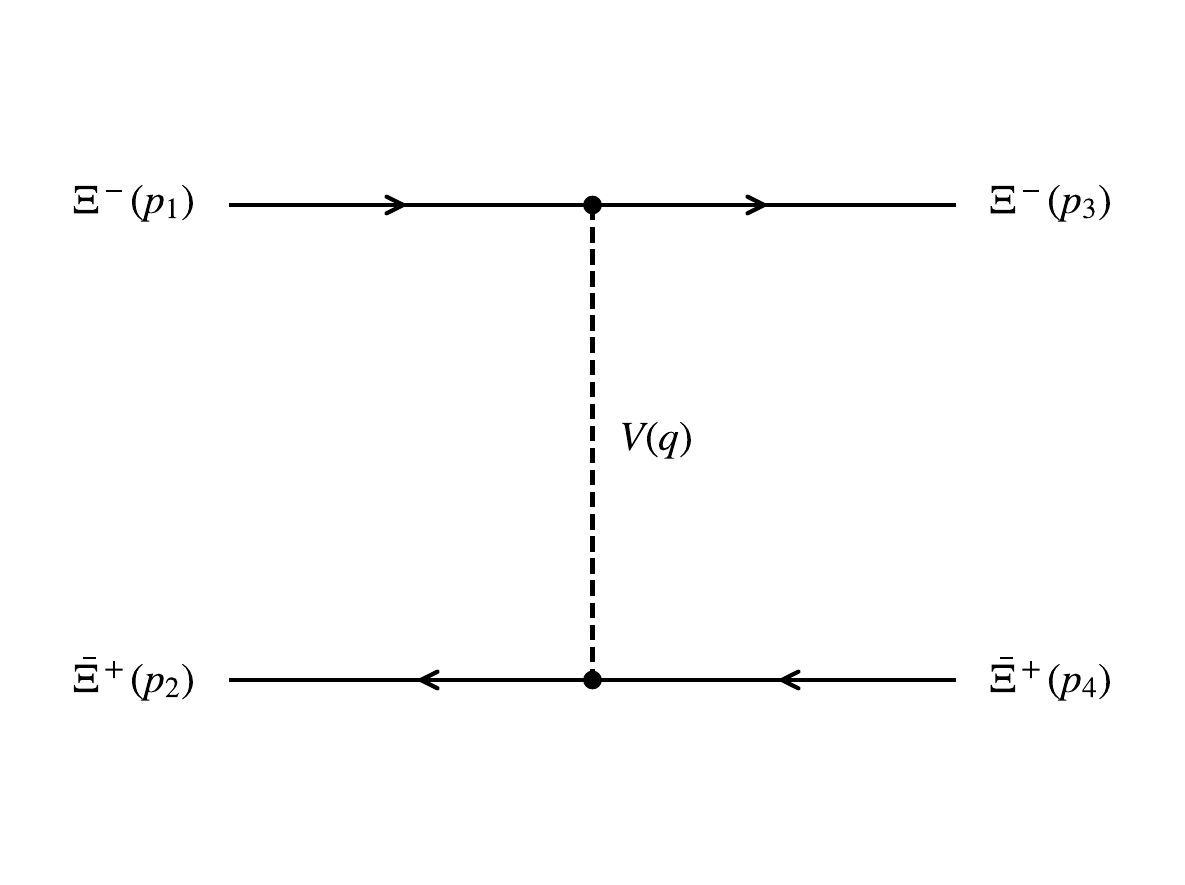}
        \caption{$V_{22}$: $\Xi^-\bar{\Xi}^+ \to \Xi^-\bar{\Xi}^+$}
        \label{fig:V22}
    \end{subfigure}
    \hfill
    \begin{subfigure}[b]{0.48\textwidth}
        \centering
        \includegraphics[width=\textwidth]{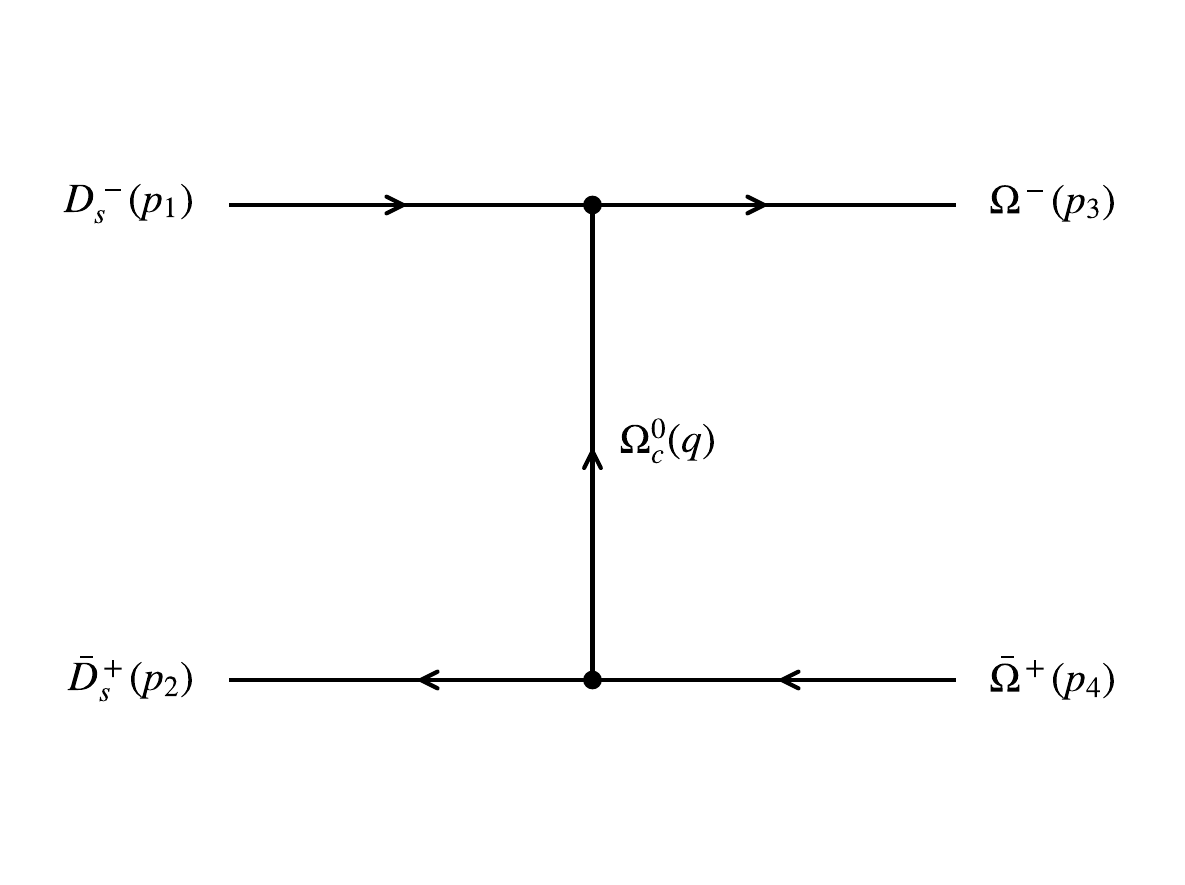}
        \caption{$V_{13}$: $D_s^-\bar{D}_s^+ \to \Omega^-\bar{\Omega}^+$}
        \label{fig:V13}
    \end{subfigure}

    \vspace{0.6cm} 

    \begin{subfigure}[b]{0.48\textwidth}
        \centering
        \includegraphics[width=\textwidth]{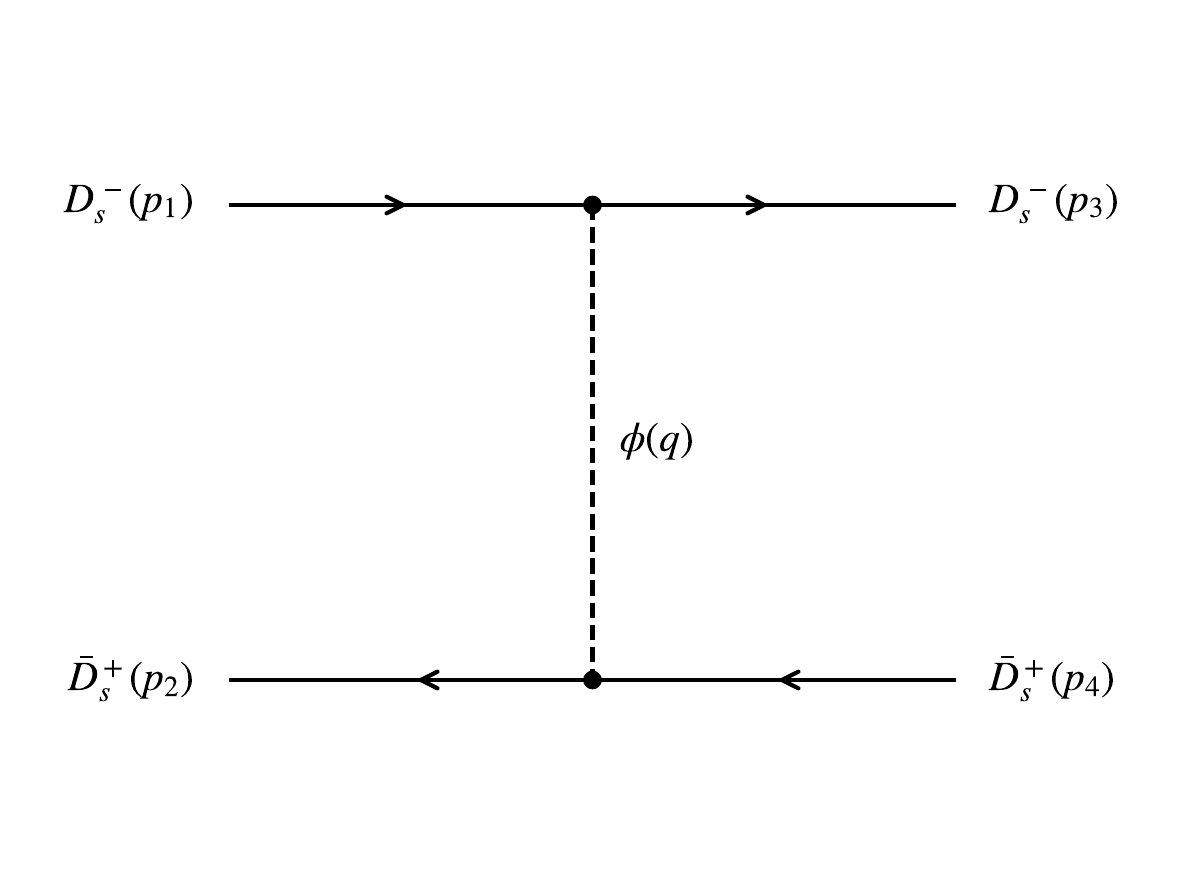}
        \caption{$V_{33}$: $D_s^-\bar{D}_s^+ \to D_s^-\bar{D}_s^+$}
        \label{fig:V33}
    \end{subfigure}
    \hfill
    \begin{subfigure}[b]{0.48\textwidth}
        \centering
        \includegraphics[width=\textwidth]{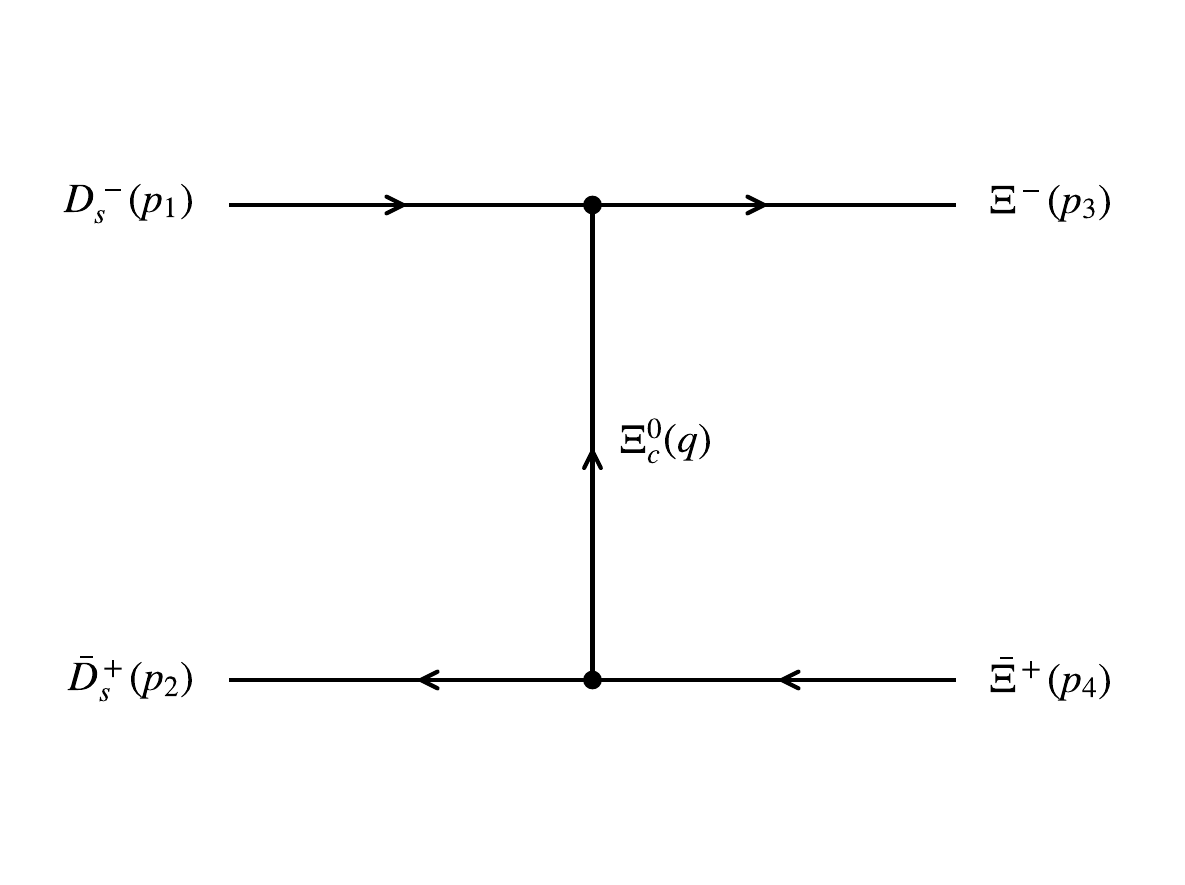}
        \caption{$V_{23}$: $D_s^-\bar{D}_s^+ \to \Xi^-\bar{\Xi}^+$}
        \label{fig:V23}
    \end{subfigure}

    \caption{Feynman diagrams driving the coupled-channel Bethe-Salpeter kernel.}
    \label{fig:all_feynman_diagrams}
\end{figure}

To accurately capture the non-perturbative threshold dynamics, \textcolor{red}{the amplitude $F(s)$ is} parameterized as a coherent superposition of $s$-channel vector resonances and the dynamically generated hadronic rescattering amplitude:
\begin{equation}
F(s)=\textcolor{red}{C_{prod}\,\frac{|\vec{p}_{D_s}|}{m_{D_s}}\,G_3(s)\,T_{13}(s)+}\sum_k A_k e^{i\phi_k}BW_k(s).
\label{eq:prod}
\end{equation}
Here, the sum over $k$ accounts for the background contributions from known charmonium or charmonium-like vector resonances (\textcolor{red}{$\psi(3770)$,} $\psi(4040)$, $Y(4230)$, $\psi(4415)$), parameterized by Breit-Wigner amplitudes $BW_k(s)\textcolor{red}{\,=1/(s-M_k^2+iM_k\Gamma_k)}$ with transition strengths $A_k$ \textcolor{red} {(in GeV$^2$)} and relative phases $\phi_k$\textcolor{red}{; the masses $M_k$ and
widths $\Gamma_k$ are fixed to the PDG values~\cite{ParticleDataGroup:2026mpi}}. The first term, $C_{prod}\textcolor{red}{\,|\vec p_{D_s}|/m_{D_s} G_3(s)}\,T_{13}(s)$, isolates the non-perturbative coupled-channel rescattering mechanism. It describes the prompt electromagnetic production of a heavy-flavor meson pair (scaled by the normalization constant $C_{prod}$), which \textcolor{red}{propagates through the $D_s^-D_s^+$ loop function $G_3(s)$ and} subsequently rescatter into the final $\Omega^-\bar{\Omega}^+$ state. \textcolor{red}{The photon transition into the $D_s^-D_s^+$ pair occurs in a
$P$ wave configuration, so the production vertex is proportional to the $D_s$ center-of-mass momentum; dividing by $m_{D_s}$ preserves a dimensionless normalization constant $C_{prod}$.} To evaluate this crucial $T$-matrix amplitude, we define a basis of three dynamically coupled channels:
\begin{itemize}
    \item Channel 1: $|\Omega^-\bar{\Omega}^+\rangle$ \textcolor{red}{($^3S_1$)}
    \item Channel 2: $|\Xi\bar{\Xi}\rangle$ \textcolor{red}{($I=0$, $^3S_1$)}
    \item Channel 3: $|D_s^-\bar{D}_s^+\rangle$ \textcolor{red}{($^1P_1$)}
\end{itemize}
\textcolor{red}{Conservation of the virtual photon quantum numbers requires all three channels to carry $J^{PC}=1^{--}$. A baryon-antibaryon system satisfies
$P=(-1)^{L+1}$ and $C=(-1)^{L+S}$, which gives the $^3S_1$ wave that we keep near threshold.
A spinless $D_s^-D_s^+$ pair obeys $P=C=(-1)^L$ yielding $J^{PC}=0^{++}$ in the $S$-wave, and therefore the photon couples to it exclusively in the $^1P_1$ wave. Every potential derived below is
projected onto these partial waves.}
The rescattering amplitudes are generated dynamically by solving the unitarized Bethe-Salpeter integral equations. The tree-level transition potentials $V_{ij}(s)$ driving this $3 \times 3$ coupled-channel system are depicted by the Feynman diagrams in Figure~\ref{fig:all_feynman_diagrams}. The left column displays the diagonal elastic scattering processes ($V_{11}, V_{22}, V_{33}$) mediated by $t$-channel vector-meson exchanges ($\phi, \omega, \rho$). The right column illustrates the off-diagonal channel-mixing transitions ($V_{12}, V_{13}, V_{23}$), which proceed through pseudoscalar-meson ($K$) and heavy-baryon ($\Omega_c^0, \Xi_c^0$) exchanges. In the following subsections, we construct the analytical form of each potential using effective Lagrangians.

\subsection{Effective Lagrangians and Transition Potentials}

To evaluate the elastic scattering within the $\Omega^- \bar{\Omega}^+$ channel (Channel 1), we consider the $t$-channel vector meson ($\phi$) exchange. Within the Local Hidden Gauge (LHG) formalism, the interaction between the spin-$3/2$ baryon decuplet ($T_\mu$) and the vector meson nonet ($V^\nu$) is governed by the Lagrangian \cite{Oset:2005ag}:
\begin{equation}
\mathcal L_{11}=\textcolor{red}{-g\langle\bar T_\mu\gamma_\nu V^\nu T^\mu\rangle},
\label{eq:L11}
\end{equation}
where $g = m_V / 2f_\pi$ is the universal vector coupling constant \cite{Oset:2010tof,Xiao:2013yca}, and $\langle \dots \rangle$ denotes the \textcolor{red}{symmetric contraction of the flavor indices of the decuplet tensor
$T^{ijk}$}. \textcolor{red}{In the near-threshold region, the baryon currents are dominated by their time components, with $\bar u_\alpha\gamma^0u^\alpha\approx
2m_\Omega$, while the spin operator reduces to the identity between the normalized $^3S_1$
states. The $q_\mu q_\nu$ term of the $\phi$ propagator vanishes due to baryon current conservation. Keeping the full momentum dependence of the propagator, the tree-level amplitude
emerges as $\mathcal M_{11}=-4C_{11}^2g^2m_\Omega^2/(m_\phi^2-t)$, with
$t=-2|\vec p_\Omega|^2(1-\cos\theta)$.}
Substituting the LHG coupling, the \textcolor{red}{$S$-wave projection, which is obtained by averaging the amplitude over the scattering angle by $V_{11}=\frac12\int_{-1}^{1}d\cos\theta\,\mathcal M_{11}$,} of the transition potential evaluates to \textcolor{red}{\cite{Mei:2022msh, Yu:2018yxl}}:
\begin{equation}
    \textcolor{red}{V_{11}(s)=-\frac{C_{11}^2\,g^2\,m_\Omega^2}{|\vec p_\Omega|^2}
\ln\!\left(1+\frac{4|\vec p_\Omega|^2}{m_\phi^2}\right)},
\label{eq:V11}
\end{equation}
where $f_\pi \approx 93$~MeV is the pion decay constant, $m_\Omega$ is the mass of the $\Omega^-$ baryon, \textcolor{red}{$m_\phi$ is the $\phi$ mass, and
$|\vec p_\Omega|=\sqrt{s/4-m_\Omega^2}$ is the center-of-mass momentum} of the interacting baryons. By evaluating the flavor tensor contraction for the $\Omega^-$ baryon ($T^{333}$), the specific isospin-flavor coefficient for the $\Omega^- \to \Omega^- + \phi$ vertex is found to be $C_{11} = 3$. 

The coupled-channel transition between the $\Xi\bar{\Xi}$ (Channel 2) and $\Omega^-\bar{\Omega}^+$ (Channel 1) states is mediated by the $t$-channel exchange of a pseudoscalar Kaon ($K$). The interaction between the spin-3/2 decuplet ($\Omega_\mu$), the spin-1/2 octet ($\Xi$), and the pseudoscalar meson is governed by the effective chiral Lagrangian \cite{Korpa:2011qg}:
\begin{equation}
    \mathcal{L}_{12} = \frac{g_{12}}{m_K} \left( \bar{\Omega}_\mu \partial^\mu K \Xi + \bar{\Xi} \partial^\mu K^\dagger \Omega_\mu \right),
\end{equation}
where $m_K$ is the Kaon mass and $g_{12}$ is the \textcolor{red}{dimensionless} coupling constant \textcolor{red}{of a single interaction vertex, such as $\Omega^-\Xi^-\bar K^0$}. Evaluating the $t$-channel tree-level Feynman diagram yields the invariant scattering amplitude:
\begin{equation}
\mathcal M_{12}=\textcolor{red}{-}\left(\frac{g_{12}}{m_K}\right)^2\frac{1}{t-m_K^2}
\left[\bar u_\mu(p_3)q^\mu u(p_1)\right]\left[\bar v(p_2)q^\nu v_\nu(p_4)\right],
\label{eq:M12}
\end{equation}
where $q\textcolor{red}{\,=p_3-p_1}$ is the transferred momentum\textcolor{red}{, $t=-|\vec q|^2$, and the overall
sign accounts for the fermion-ordering sign along the antibaryon line.} Because the exchanged particle is a spin-0 pseudoscalar, the internal propagator reduces strictly to a scalar function $\Delta_F(q) \propto (t - m_K^2)^{-1}$, devoid of additional tensor structure.
In the near-threshold heavy baryon limit, the Dirac spinors for the spin-1/2 $\Xi$ baryons \textcolor{red}{reduce to $\sqrt{2m_\Xi}$ times two-component Pauli spinors, while the} Rarita-Schwinger vector-spinors \textcolor{red}{reduce to} spatial polarization vectors $\vec{u}$ and $\vec{v}$ \textcolor{red}{coupled to Pauli spinors through the Clebsch-Gordan coefficients}. The spinor contractions thus reduce to purely spatial inner products, $(\vec{q} \cdot \vec{u})(\vec{q} \cdot \vec{v})$\textcolor{red}{, multiplied by
the normalization factor $4m_\Omega m_\Xi$}.

We project the interaction onto the isoscalar ($I=0$) basis dictated by the initial and final states. While the $\Omega^-$ baryon is an inherent isosinglet, the $\Xi$ baryon forms an isodoublet ($\Xi^-, \Xi^0$). Assuming exact $SU(2)$ isospin symmetry ($m_{K^0} \approx m_{K^+}$), the $t$-channel transition amplitudes for the charged and neutral paths are identical. Summing over these mediating paths yields an overall isospin projection factor of $ \sqrt{2}$.
To project the amplitude onto the spherically symmetric $S$-wave ($L=0$) state, we evaluate the angular average of the spatial tensor over the solid angle. Exploiting rotational invariance, the tensor product reduces isotropically as $q_i q_j \to \frac{1}{3}|\vec{q}|^2 \delta_{ij}$. \textcolor{red}{Combining this factor of $1/3$ with the coupling of the $\Xi\bar\Xi$ and
$\Omega^-\bar\Omega^+$ spins to $S=1$ results in an overall spin factor of $\sqrt{10}/9$.} 
\textcolor{red}{Taking into account the isospin factor $\sqrt2$, the normalization $4m_\Omega m_\Xi$, and the
spin factor $\sqrt{10}/9$,} the integration over the scattering angle $\theta$ yields the
analytical $S$-wave potential:
\begin{equation}
\textcolor{red}{V_{12}(s)=\frac{8\sqrt5}{9}\,\frac{m_\Omega m_\Xi}{m_K^2}\,g_{12}^2
\left[1+\frac{m_K^2}{4pp'}\ln\left(\frac{A_{12}+2pp'}{A_{12}-2pp'}\right)\right]},
\label{eq:V12}
\end{equation}
where $p = |\vec{p}_\Xi|$ and $p' = |\vec{p}_\Omega|$ denote the center-of-mass spatial momenta of the initial and final states. The kinematic, angle-independent momentum transfer parameter is defined as $A_{12} = (E_\Xi - E_\Omega)^2 - p^2 - p'^2 - m_K^2$.

To evaluate the transition potential for the $D_s \bar{D}_s \to \Omega^- \bar{\Omega}^+$ process via $t$-channel $\Omega_c$ exchange, we employ the standard effective interaction Lagrangian \cite{Korpa:2011qg},
\begin{equation}
    \mathcal{L}_{13} = \frac{g_{13}}{m_{D_s}} \left( \bar{\Omega}_\mu \partial^\mu D_s \Omega_c + \bar{\Omega}_c \partial^\mu D_s^\dagger \Omega_\mu \right)  \label{eq:L13}
\end{equation}
The corresponding invariant Feynman amplitude is derived as
\begin{equation}
    \mathcal{M}_{13} =  \left( \frac{g_{13}}{m_{D_s}} \right)^2 \frac{1}{t - M_{\Omega_c}^2} \left[ \bar{u}_\mu(p_3) q^\mu (\slashed{q} + M_{\Omega_c}) q^\nu v_\nu(p_4) \right].
\end{equation}
Because we focus on the near-threshold region where the spatial momenta are parametrically small ($|\vec{p}| \ll M_{\Omega_c}$), we apply the non-relativistic reduction to the heavy Rarita-Schwinger spinors, utilizing the transversality constraint $p^\mu u_\mu = 0$\textcolor{red}{,
and expand them to their leading nonzero momentum contributions, keeping all four spin
projections of $\Omega$ and $\bar\Omega$, as well as the recoil term proportional to
$M_{\Omega_c}/m_\Omega$}. The tensor structure reduces to a purely spatial inner product. \textcolor{red}{The initial $D_s\bar D_s$ state is in the $^1P_1$ configuration, while the final
$\Omega^-\bar\Omega^+$ state is in the $^3S_1$ wave. The $\Omega$ and $\bar\Omega$ spins are coupled to
$S=1$. Choosing the initial momentum along $z$ with $M_J=0$, the partial wave projection is performed using
$V_{13}=\frac{1}{2\sqrt3}\int_{-1}^{1}d\cos\theta\,\mathcal M_{13}$, where the factor $1/\sqrt3$
arises from the normalization of $Y_{10}$. The external normalization prefactor
$2m_\Omega$, the spin factor $2/(3\sqrt5)$ and the orbital projection factor give the prefactor shown below.}
Evaluating the remaining solid-angle projection over the $t$-channel propagator yields the analytical \textcolor{red}{$P\to S$} transition potential:
\begin{equation}
   \textcolor{red}{V_{13}(s)=\frac{4m_\Omega\,g_{13}^2}{3\sqrt{15}\,m_{D_s}^2}
\Big[f_0\,I_0+f_1\,I_1+f_2\,I_2\Big]},
\label{eq:V13}
\end{equation}
where $p = |\vec{p}_{D_s}|$ and $p' = |\vec{p}_\Omega|$ are the center-of-mass three-momenta of the initial and final states, and the kinematic parameter $A$ is defined as $A = (E_{D_s} - E_\Omega)^2 - p^2 - p'^2 - M_{\Omega_c}^2$\textcolor{red}{$=-(p^2+p'^2+M_{\Omega_c}^2)$}. \textcolor{red}{With $\rho=M_{\Omega_c}/m_\Omega$, the coefficients in the numerator become}
\begin{equation}
\textcolor{red}{f_0=p\left[-p^2+(\rho-1)\,p'^2\right],
f_1=p'\left[(3+\rho)\,p^2+(1+\rho)\,p'^2\right],
f_2=-(2+3\rho)\,p\,p'^2,}
\end{equation}
\textcolor{red}{and the angular integrals over the propagator, $1/(M_{\Omega_c}^2-t)=-1/(A+2pp'z)$ are}
\begin{equation}
\textcolor{red}{I_n=\frac12\int_{-1}^{1}\frac{z^n\,dz}{-A-2pp'z},
I_0=\frac{1}{4pp'}\ln\left(\frac{A-2pp'}{A+2pp'}\right),
I_1=-\frac{1+A\,I_0}{2pp'},
I_2=-\frac{A}{2pp'}\,I_1 .}
\label{eq:In}
\end{equation}
\textcolor{red}{with $z=\cos\theta$. The isospin factor equals one. As expected for an initial $P$ wave state, $V_{13}$ vanishes linearly in $p$ at the $D_s\bar D_s$ threshold. This momentum dependence is kept in the
interaction kernel and is not absorbed into $g_{13}$.}

For the elastic $\Xi\bar{\Xi}$ scattering (Channel 2), the interaction is dominated by the $t$-channel exchange of light vector mesons ($V = \rho, \omega, \phi$). The interaction between the spin-$1/2$ baryon octet ($B$) and the vector meson nonet ($V^\mu$) is governed by the Lagrangian\textcolor{red}{~\cite{Oset:2010tof}}:
\begin{equation}
    \mathcal L_{22}=g\textcolor{red}{\Big(}\langle\bar B\gamma_\mu[V^\mu,B]\rangle
\textcolor{red}{+\langle\bar B\gamma_\mu B\rangle\langle V^\mu\rangle\Big),\qquad
\langle V^\mu\rangle=\sqrt2\,\omega^\mu+\phi^\mu},
\end{equation}
where $g = m_V / 2f_\pi$ is the universal vector coupling constant, and $\langle \dots \rangle$ denotes the trace over SU(3) flavor space.
We focus on the isosinglet state ($I=0$), which dictates that the physical channel is constructed as a coherent superposition, $|\Xi\bar{\Xi}\rangle_{I=0} = \frac{1}{\sqrt{2}} \left( |\Xi^-\bar{\Xi}^+\rangle + |\Xi^0\bar{\Xi}^0\rangle \right)$. The total transition amplitude requires evaluating the flavor traces across four \textcolor{red}{charge-channel amplitudes}, comprising both elastic transitions and charge-exchange processes ($\Xi^-\bar{\Xi}^+ \leftrightarrow \Xi^0\bar{\Xi}^0$). \textcolor{red}{Summing
the products of the vector-meson couplings results in the exchange weights $C_\rho=3/2$,
$C_\omega=1/2$ and $C_\phi=4$, which add up to $C_{22}=6$ for a common vector mass. Using the
same static current approximation as for $V_{11}$, the amplitude takes the form
$\mathcal M_{22}=-4m_\Xi^2g^2\sum_V C_V/(m_V^2-t)$.} Projecting the invariant amplitude onto the $S$-wave state, the potential becomes \cite{Yu:2018yxl,Lyu:2026rsm, Mei:2022msh}:
\begin{equation}
\textcolor{red}{V_{22}(s)=-\frac{g^2m_\Xi^2}{|\vec p_\Xi|^2}\sum_{V=\rho,\omega,\phi}C_V
\ln\!\left(1+\frac{4|\vec p_\Xi|^2}{m_V^2}\right)},
\label{eq:V22}
\end{equation}
To preserve rigorous $SU(2)$ isospin symmetry within the interacting channel, we adopt the isospin-averaged baryon mass $m_\Xi$, \textcolor{red}{and $|\vec p_\Xi|=\sqrt{s/4-m_\Xi^2}$ denotes the center-of-mass momentum} of the $\Xi$ baryons.

The coupled-channel transition between the initial $D_s\bar{D}_s$ state (Channel 3) and the $\Xi\bar{\Xi}$ state (Channel 2) is governed by the $t$-channel exchange of a heavy $\Xi_c$ baryon. Following the chiral Lagrangian framework for pseudoscalar-baryon interactions \cite{Yan:1992gz}, the vertex is described by the pseudo-scalar coupling:
\begin{equation}
    \mathcal{L}_{23} = i g_{23} \left( \bar{\Xi} \gamma_5 D_s \Xi_c + \bar{\Xi}_c \gamma_5 D_s^\dagger \Xi \right). 
\end{equation}
Evaluating the corresponding tree-level Feynman diagram yields the invariant scattering amplitude:
\begin{equation}
    \mathcal{M}_{23} = \textcolor{red}{-} g_{23}^2 \frac{1}{t - M_{\Xi_c}^2} \left[ \bar{u}(p_3) \gamma_5 (\slashed{q} + M_{\Xi_c}) \gamma_5 v(p_4) \right].
\end{equation}
Exploiting the anti-commutation relations of the Dirac algebra ($\gamma_5 \slashed{q} \gamma_5 = -\slashed{q}$), and utilizing the on-shell Dirac equation $\bar{u}(p_3)\slashed{p}_3 = m_\Xi \bar{u}(p_3)$, the numerator reduces to the form $\bar{u}(p_3) (\slashed{p}_1 + M_{\Xi_c} - m_\Xi) v(p_4)$.
In the near-threshold limit, the spatial momenta are parametrically small, \textcolor{red}{and the leading spinors yield two first-order terms
in the momenta. While the temporal energy term ($E_{D_s}\bar u\gamma^0v$) cancels
between the small  components of the spinors, they add in the scalar overlap, giving
$\bar u(p_3)v(p_4)\approx-2\,\vec p_3\cdot(\chi^\dagger\vec\sigma\eta)$, which remains nonzero. Thus, the
transition is  driven by the spatial momentum component and the scalar mass term}:
\begin{equation}
\textcolor{red}{\bar u(p_3)(\slashed p_1+M_{\Xi_c}-m_\Xi)v(p_4)\approx
-2\left[m_\Xi\,\vec p_1+(M_{\Xi_c}-m_\Xi)\,\vec p_3\right]\cdot(\chi^\dagger\vec\sigma\eta)}.
\label{eq:N23}
\end{equation}
This shows that the $\gamma_5$ parity constraint naturally selects the $L=1$ ($P$-wave) spatial configuration for the initial $D_s\bar{D}_s$ pair, fulfilling the $J^{PC} = 1^{--}$ quantum numbers dictated by the $e^+e^-$ annihilation process.
Similar to the $V_{12}$ transition, the final state requires a projection onto the $I=0$ isosinglet basis. The transition from the isosinglet $D_s\bar{D}_s$ pair to the isodoublet $\Xi\bar{\Xi}$ superposition introduces an isospin Clebsch-Gordan factor of $\sqrt{2}$. \textcolor{red}{Projecting onto the $\Xi\bar\Xi$ $^3S_1$ state yields a spin factor of $\sqrt2$, and applying the $P\to S$ partial wave projection as in $V_{13}$,
$V_{23}=\frac{1}{2\sqrt3}\int_{-1}^{1}d\cos\theta\,\mathcal M_{23}$,} the integration over the scalar $t$-channel denominator evaluates exactly to the analytical potential:
\begin{equation}
\textcolor{red}{V_{23}(s)=-\frac{2g_{23}^2}{\sqrt3}\left[\left(\frac{m_\Xi}{2p'}
-\frac{(M_{\Xi_c}-m_\Xi)\,A_{23}}{4p^2p'}\right)\ln\left(\frac{A_{23}+2pp'}{A_{23}-2pp'}\right)
+\frac{M_{\Xi_c}-m_\Xi}{p}\right]},
\label{eq:V23}
\end{equation}
where $p = |\vec{p}_{D_s}|$ and $p' = |\vec{p}_\Xi|$ are the center-of-mass momenta, and the angle-independent kinematic parameter is defined as $A_{23} = (E_{D_s} - E_\Xi)^2 - p^2 - p'^2 - M_{\Xi_c}^2$.

For the elastic $D_s^-\bar{D}_s^+$ scattering (Channel 3), the interaction is dominated by the $t$-channel exchange of the isoscalar vector meson $\phi(s\bar{s})$.  The interaction between pseudoscalar mesons ($P$) and vector mesons ($V^\mu$) is governed by the Lagrangian\textcolor{red}{, with the flavor matrices are generalized to include charmed
mesons~\cite{Bayar:2022dqa}}:
\begin{equation}
\mathcal{L}_{VPP} = -i g \langle V^\mu [P, \partial_\mu P] \rangle,
\label{eq:L_VPP}
\end{equation}
where $g = m_V / (2f_\pi)$ is the universal vector coupling constant, $f_\pi \approx 93$~MeV is the pion decay constant, and $\langle \dots \rangle$ denotes the trace over flavor space.
Evaluating the tree-level $t$-channel Feynman diagram \textcolor{red}{leads to
$\mathcal M_{33}=C_{33}\,g^2\,(p_1+p_3)\cdot(p_2+p_4)/(t-m_\phi^2)$, where the $q_\mu q_\nu$ term of
the $\phi$ propagator vanishes since $q\cdot(p_1+p_3)=0$ in the equal mass case. Following the same procedure used for the
baryon currents, we keep the static approximation for the numerator,
$(p_1+p_3)\cdot(p_2+p_4)\to4m_{D_s}^2$. Because the $D_s^-D_s^+$ pair is in the $^1P_1$ partial wave, we
project} the invariant scattering amplitude onto the \textcolor{red}{$P$-wave ($L=1$) configuration,
$V_{33}=\frac12\int_{-1}^{1}d\cos\theta\,\cos\theta\,\mathcal M_{33}$, which} yields the analytical elastic potential:
\begin{equation}
\textcolor{red}{V_{33}(s)=-\frac{C_{33}\,g^2\,m_{D_s}^2}{2|\vec p_{D_s}|^4}
\left[\left(m_\phi^2+2|\vec p_{D_s}|^2\right)\ln\!\left(1+\frac{4|\vec p_{D_s}|^2}{m_\phi^2}\right)
-4|\vec p_{D_s}|^2\right]},
\label{eq:V33}
\end{equation}
where $m_{D_s}$ is the mass of the $D_s^\pm$ meson and $s$ is the center-of-mass energy squared. Evaluating the flavor trace for the $D_s^-\bar{D}_s^+$ isoscalar pair interacting via $\phi(s\bar{s})$ exchange yields the  coefficient $C_{33} = 1$. \textcolor{red}{Close to threshold,
this reduces to $V_{33}\simeq-8C_{33}g^2m_{D_s}^2|\vec p_{D_s}|^2/(3m_\phi^4)$, exhibiting the $|\vec p|^2$ behavior of an
elastic $P$ wave interaction.}

\textcolor{red}{For energies below the threshold of channel $i$, the squared momentum, $|\vec p_i|^2=s/4-m_i^2$, becomes negative, causing
the logarithms in Eqs.~(\ref{eq:V11}), (\ref{eq:V22}) and (\ref{eq:V33}) to generate the left-hand
cut associated with the $t$-channel exchange, beginning at $|\vec p_i|^2=-m_V^2/4$. In the case of $\phi$ exchange in
the $D_s^-D_s^+$ channel, this branch point is located at $\sqrt s=3.80$~GeV,
which falls within the measured energy range. As the on-shell factorization cannot handle this cut consistently and can generate artifacts close to the branch
point \cite{Gulmez:2016scm}, for $\sqrt{s} < 2 m_i$ we replace the diagonal kernels by their threshold
expansions to first order in $|\vec p_i|^2$ \cite{Epelbaum:2008ga},}
\begin{align}
\textcolor{red}{V_{11}}&\textcolor{red}{\to-\frac{4C_{11}^2g^2m_\Omega^2}{m_\phi^2}
\left(1-\frac{2|\vec p_\Omega|^2}{m_\phi^2}\right),\qquad
V_{22}\to-4g^2m_\Xi^2\sum_V\frac{C_V}{m_V^2}\left(1-\frac{2|\vec p_\Xi|^2}{m_V^2}\right),}
\nonumber\\
\textcolor{red}{V_{33}}&\textcolor{red}{\to-\frac{8C_{33}\,g^2m_{D_s}^2}{3m_\phi^4}\,|\vec p_{D_s}|^2,}
\label{eq:Vthr}
\end{align}
\textcolor{red}{which match the full expressions at threshold in both value and slope with respect to
$|\vec p_i|^2$. Within the energy range considered in the fit, this modifies only $V_{33}$, at energies $\sqrt s<2m_{D_s}$.}
\subsection{Coupled-Channel 
\label{sec:unitarization}
Unitarization and Loop Regularization}

    To dynamically generate the molecular resonances and evaluate the physical scattering observables, the tree-level transition potentials $V_{ij}$ must be unitarized. We achieve this by solving the Bethe-Salpeter equation in the on-shell factorization approximation. Because the driving potentials \textcolor{red}{are evaluated on-shell within the partial waves of the three
channels}, they can be factorized outside of the loop momentum integral. This algebraically reduces the coupled-channel integral equations to a strict matrix inversion:
\begin{equation}
    T(s) = [1 - V(s)G(s)]^{-1} V(s),
    \label{eq:T}
\end{equation}
where $V(s)$ is the $3 \times 3$ potential matrix derived in the previous section, $T(s)$ is the unitarized scattering amplitude matrix, and $G(s) = \text{diag}(G_1, G_2, G_3)$ is the diagonal matrix of the two-body loop functions for the $\Omega^-\bar{\Omega}^+$, $\Xi\bar{\Xi}$, and $D_s\bar{D}_s$ channels, respectively. 
The two-body loop function for the $i$-th channel is defined via the momentum integral \cite{Gamermann:2009uq, Montana:2022inz}:
\begin{equation}
    G_i(s) = i \int \frac{d^4q}{(2\pi)^4} \frac{1}{(q^2 - m_{1i}^2 + i\epsilon)} \frac{1}{((P-q)^2 - m_{2i}^2 + i\epsilon)},
    \label{eq:G}
\end{equation}
where $P$ is the total four-momentum of the system, and $m_{1i}, m_{2i}$ are the masses of the constituents in channel $i$. 
This integral is logarithmically divergent and requires regularization. Following the standard chiral unitary approach \cite{Feijoo:2022rxf, Feijoo:2024bvn,Li:2026umb,Lyu:2026rsm,Wang:2025jcq,Wang:2024fsz,Nishibuchi:2023acl,Rahmani:2026cjb}, we regulate the loop function by imposing a three-momentum cutoff, $q_{max}$, in the center-of-mass frame. Integrating over the temporal component analytically via contour integration, and integrating the spatial momentum up to $|\vec{q}| = q_{max}$, yields the exact analytical expression for the regularized loop function \cite{Rahmani:2026gcx,Rahmani:2025uut}:
 \begin{align}
G_i^{\textcolor{red}{cut}}(s)=\frac{1}{16\pi^2s}\bigg\{&\sigma\left[\arctan\left(\frac{s+\Delta}{\sigma\lambda_1}\right)
+\arctan\left(\frac{s-\Delta}{\sigma\lambda_2}\right)\right]
\nonumber\\
&-(s+\Delta)\ln\left(\frac{q_{max}}{m_{1i}}(1+\lambda_1)\right)
-(s-\Delta)\ln\left(\frac{q_{max}}{m_{2i}}(1+\lambda_2)\right)\bigg\},
\label{eq:Gcut}
\end{align}
where $\Delta = m_{1i}^2 - m_{2i}^2$, $\lambda_k = \sqrt{1 + m_{ki}^2 / q_{max}^2}$, and $\sigma=\sqrt{\textcolor{red}{-}(s-(m_{1i}+m_{2i})^2)(s-(m_{1i}-m_{2i})^2)}$.

\textcolor{red}{Under a sharp cutoff, however, the on-shell singularity of Eq. (\ref{eq:G}) lies
outside the integration domain whenever the center-of-mass momentum $|\vec p_i|$ exceeds $q_{max}$. Consequently,
$G_i^{cut}$ becomes real above $\sqrt s=2\sqrt{m_i^2+q_{max}^2}$, and the open channels would lose their absorptive parts within the energy
range of interest. To avoid this artifact, we compute Eq. (\ref{eq:G}) using dimensional regularization, which is
equivalent to applying a once-subtracted dispersion relation \cite{Oller:2000fj,Guo:2005wp}. The resulting loop function depends on a subtraction
constant $a_i(\mu)$ at the regularization scale $\mu$~\cite{Oller:2000fj,Guo:2005wp} and yields
${\rm Im}\,G_i(s)=-|\vec p_i|/(8\pi\sqrt s)$ at all energies above
threshold, thereby ensuring that $T(s)$ fulfills two-body unitarity across every open
channel. Following Ref.~\cite{Montana:2022inz}, we determine the subtraction constants by
equating $G_i=G_i^{cut}$ at threshold ($s=4m_i^2$). The matched loop function is independent of
$\mu$, with the cutoff serving solely to set the real part of $G_i$ at threshold. Since both $G_i$ and the
potentials $V_{ij}$ are dimensionless, $T(s)$ and $F(s)$ are dimensionless, allowing
Eq. (\ref{eq:T}) to couple all three channels under a unified normalization.}

In our numerical evaluations, the cutoff parameter is fixed to $q_{max} = 0.750$~GeV \cite{Li:2023olv}, a characteristic hadronic scale widely successful in reproducing the properties of dynamically generated baryon and meson resonances. The resulting physical cross section for the $e^+e^- \to \Omega^-\bar{\Omega}^+$ process is then \textcolor{red}{determined by Eqs.(\ref{eq:sigma})
and (\ref{eq:prod}), in which $T_{13}(s)$ incorporates} all non-perturbative rescattering mechanisms.

\section{Numerical Analysis and Deep Learning Results}
\label{sec:results}
In this section, we present the numerical analysis of our $3 \times 3$ coupled-channel framework ($|\Omega^-\bar{\Omega}^+\rangle$, $|\Xi\bar{\Xi}\rangle$, $|D_s^-\bar{D}_s^+\rangle$) to analyze the $e^+e^- \to \Omega^-\bar{\Omega}^+$ Born cross section measured by the BESIII collaboration \cite{BESIII:2025fph}. The process is executed in three phases: 
\begin{enumerate}
    \item A line-shape fit to find  \textcolor{red}{the best-fit} couplings;
    \item A deep learning approach to map out all possible combinations of couplings that can explain the BESIII data\textcolor{red}{, treating the charmonium parameters as nuisance parameters and benchmarking the results against nested sampling under the same likelihood};
    \item Pole extraction via a Cauchy-Riemann Physics-Informed Neural Network\textcolor{red}{, and verified by directly searching for the zeros of $\det[1-V(E)G^{(\rm sheet)}(E)]$ across the corresponding Riemann sheets}.
\end{enumerate}
\subsection{Phenomenological Line-Shape}
\label{subsec:minuit_fit}
We first perform a $\chi^2$ minimization of the $e^+e^- \to \Omega^-\bar{\Omega}^+$ Born cross section using the unitarized \textcolor{red}{ coupled-channel $J^{PC}=1^{--}$} Bethe-Salpeter equation ($T = [1-VG]^{-1}V$) defined in Eq.~\eqref{eq:T}. The \textcolor{red}{subtraction
constants entering the loop functions $G_i(s)$ are determined by matching onto the cutoff-regularized loop integral at each respective threshold
(see Sec.~\ref{sec:unitarization}), using} $q_{max} = 0.750$~GeV, a hadronic scale consistent with Refs. \cite{Yu:2018yxl,Xiao:2024kfq}.
The fitting parameters comprise the prompt electromagnetic production normalization ($C_{prod}$), the three effective transition couplings ($g_{13}, g_{12}, g_{23}$), and the amplitudes ($A_k$) and phases ($\phi_k$) of the background charmonium vector resonances (\textcolor{red}{$\psi(3770)$,} $\psi(4040)$, $Y(4230)$, $\psi(4415)$). 

\textcolor{red}{Our analysis incorporates all 34 energy data points in Ref. \cite{BESIII:2025fph}, including the eight points where
$\sigma_B=0$. At each energy point, the total uncertainty $\delta_i$ is obtained by averaging the upper and lower statistical errors and combining the result in quadrature
with the systematic uncertainty. We then minimize the function}
\begin{equation}
\textcolor{red}{\chi^2=\sum_{i=1}^{34}\left(\frac{\sigma_i^{exp}-\sigma^{th}(\sqrt{s_i})}{\delta_i}\right)^2,}
\label{eq:chi2}
\end{equation}

\begin{table}[htbp]
\centering
\caption{Best-fit parameters extracted from the $e^+e^- \to \Omega^-\bar{\Omega}^+$ Born cross section. The fit achieves a reduced \textcolor{red}{$\chi^2/\text{d.o.f.} = 0.803$}.}
\begin{tabular}{lc}
\hline\hline
Parameter & Value \\
\hline
$C_{prod}$ &  \textcolor{red}{$15.081 \pm 38.693$ }\\
$g_{13}$ & \textcolor{red}{$2.490^{+3.510}_{-2.490}$ } \\
$g_{12}$ &  \textcolor{red}{$4.685^{+1.537}_{-0.367}$ }\\
$g_{23}$ &  \textcolor{red}{$0.772  \pm 1.466$ } \\
\hline
\textcolor{red}{$A_{\psi(3770)} (GeV^2)$, $\phi_{\psi(3770)} (rad)$ }  & \textcolor{red}{$0.002^{+0.000}_{-0.000}$, $-1.112  \pm 2.920$ }
\\
$A_{\psi(4040)}$ \textcolor{red}{$(GeV^2)$}, $\phi_{\psi(4040)}$ \textcolor{red}{$(rad)$} & \textcolor{red}{$0.003^{+0.001}_{-0.003}$, $ -2.267^{+2.292}_{-1.527}$ } \\
$A_{Y(4230)}$ \textcolor{red}{$(GeV^2)$}, $\phi_{Y(4230)}$ \textcolor{red}{$(rad)$}&  \textcolor{red}{$0.003^{+0.000}_{-0.000}$, $2.426^{+1.297}_{-1.247}$ } \\
$A_{\psi(4415)}$ \textcolor{red}{$(GeV^2)$}, $\phi_{\psi(4415)}$\textcolor{red}{$(rad)$} &  \textcolor{red}{$ 0.002^{+0.001}_{-0.001}$, $1.121^{+1.098}_{-1.223}$} \\
\hline
\textcolor{red}{$C_{prod} g_{13}^2$} &  \textcolor{red}{$93.482^{+423.642}_{-89.114}$ }\\
\textcolor{red}{$g_{23}/g_{13}$} & \textcolor{red}{$0.310^{+0.493}_{-0.082}$ }\\
\hline\hline
\end{tabular}
\label{tab:minuit_results}
\end{table}

\begin{figure}[htbp]
    \centering
    \includegraphics[width=0.82\textwidth]{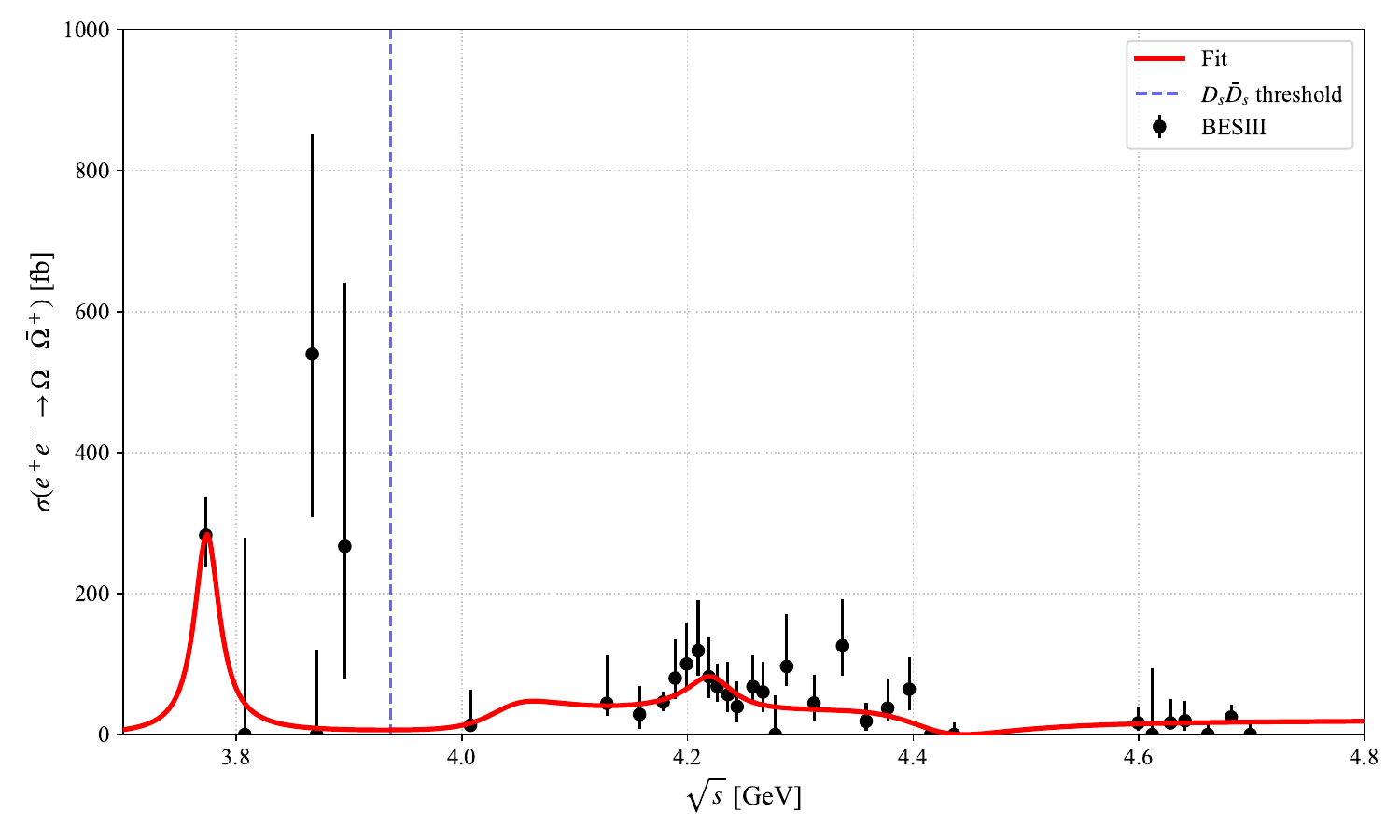}
    \caption{The $e^+e^- \to \Omega^-\bar{\Omega}^+$ Born cross section measured by BESIII \cite{BESIII:2025fph} compared with the theoretical line-shape fit. The vertical blue dashed line denotes the $D_s^-\bar{D}_s^+$ threshold.}
    \label{fig:minuit_fit}
\end{figure}

The extracted parameters and error bounds are summarized in Table~\ref{tab:minuit_results}, and the corresponding line-shape fit is illustrated in Figure~\ref{fig:minuit_fit}. The fit achieves an optimal reduced $\chi^2/\text{d.o.f.} = \textcolor{red}{17.66/22=0.80}$\textcolor{red}{, corresponding to a $p$-value of
0.73}, demonstrating good agreement with the experimental data across the energy range. \textcolor{red}{Every one of the 34 data points falls within $2\sigma$ of the curve, with the largest deviation
at $\sqrt s=3.867$ GeV, just below the $D_s\bar D_s$ threshold.}

\textcolor{red}{Nevertheless, the fit does not uniquely constrain every coupling. The multi-start minimization identifies ten distinct minima with $\chi^2$ values agreeing to within $10^{-4}$, across which $C_{prod}$ spans from 2.6 to 74,
$g_{13}$ between 1.1 and its upper bound of 6, and $g_{23}$ between 0.35 and 1.86. Because $T_{13}$ is, to
a very good approximation, linear in both $V_{13}\propto g_{13}^2$ and $V_{23}\propto g_{23}^2$, the
production amplitude depends on $C_{prod}$, $g_{13}$ and $g_{23}$ only through two combinations, which
stay fixed along this flat valley:}
\begin{equation}
\textcolor{red}{C_{prod}\,g_{13}^2=93^{+424}_{-89},\qquad g_{23}/g_{13}=0.31^{+0.49}_{-0.08}.}
\label{eq:comb}
\end{equation}
\textcolor{red}{The
magnitudes of $g_{13}$ and $g_{23}$ are not fixed by the
BESIII data.}
 Our extracted Kaon-exchange coupling, $g_{12}=\textcolor{red}{4.69^{+1.54}_{-0.37}}$, shows an agreement with theoretical predictions. In particular, within the general framework of the chiral quark-soliton model, Yang and Kim \cite{Yang:2018idi} predicted the pseudovector coupling constant for the $K \Xi \Omega$ vertex to be $ 8.130 \pm 0.080$ when accounting for SU(3) flavor symmetry breaking (and $8.339 \pm 0.079$ in the exact SU(3) limit). \textcolor{red}{Because their coupling includes an SU(3) factor of $\sqrt2$ relative to our normalization, dividing by $\sqrt2$ yields
$5.75\pm0.06$ and $5.90\pm0.06$, both of which fall within our interval.} The consistency between our phenomenologically extracted value from the BESIII cross-section data and the independent chiral quark-soliton prediction \textcolor{red}{favors} the magnitude of our $t$-channel Kaon-exchange transition potential. \textcolor{red}{In Sec.~\ref{subsec:bayesian_mdn} we compute the full
posterior distribution for $C_{prod}$ and the three couplings, treating the charmonium parameters
as nuisance parameters.}

\subsection{Uncertainty Quantification and Degeneracy Mapping}
\label{subsec:bayesian_mdn}

In coupled-channel frameworks, interference between prompt production and rescattering mechanisms can introduce non-linear parameter degeneracies and multi-modal probability landscapes. \textcolor{red}{Section \ref{subsec:minuit_fit} illustrates such a degeneracy: the couplings reside along a
flat valley of the $\chi^2$, while MINUIT provides only local uncertainties around a single minimum.} To map the complete parameter space and quantify theoretical uncertainties, we employ SBI with a Mixture Density Network (MDN)~\cite{Cranmer:2019eaq,Burton:2021tsd}. This deep learning model learns the full probability distribution 
across our four-dimensional parameter space $C_{prod}, g_{13},
g_{12}, g_{23}$, \textcolor{red}{with the charmonium parameters integrated out as nuisance parameters, thereby mapping out
the parameter degeneracies that a local fit fails to capture} \cite{Albert:2024zsh}. \textcolor{red}{We cross check the
MDN results against nested sampling using the same likelihood and prior distributions.} 

\begin{figure}[htbp]
    \centering
    \includegraphics[width=0.85\textwidth]{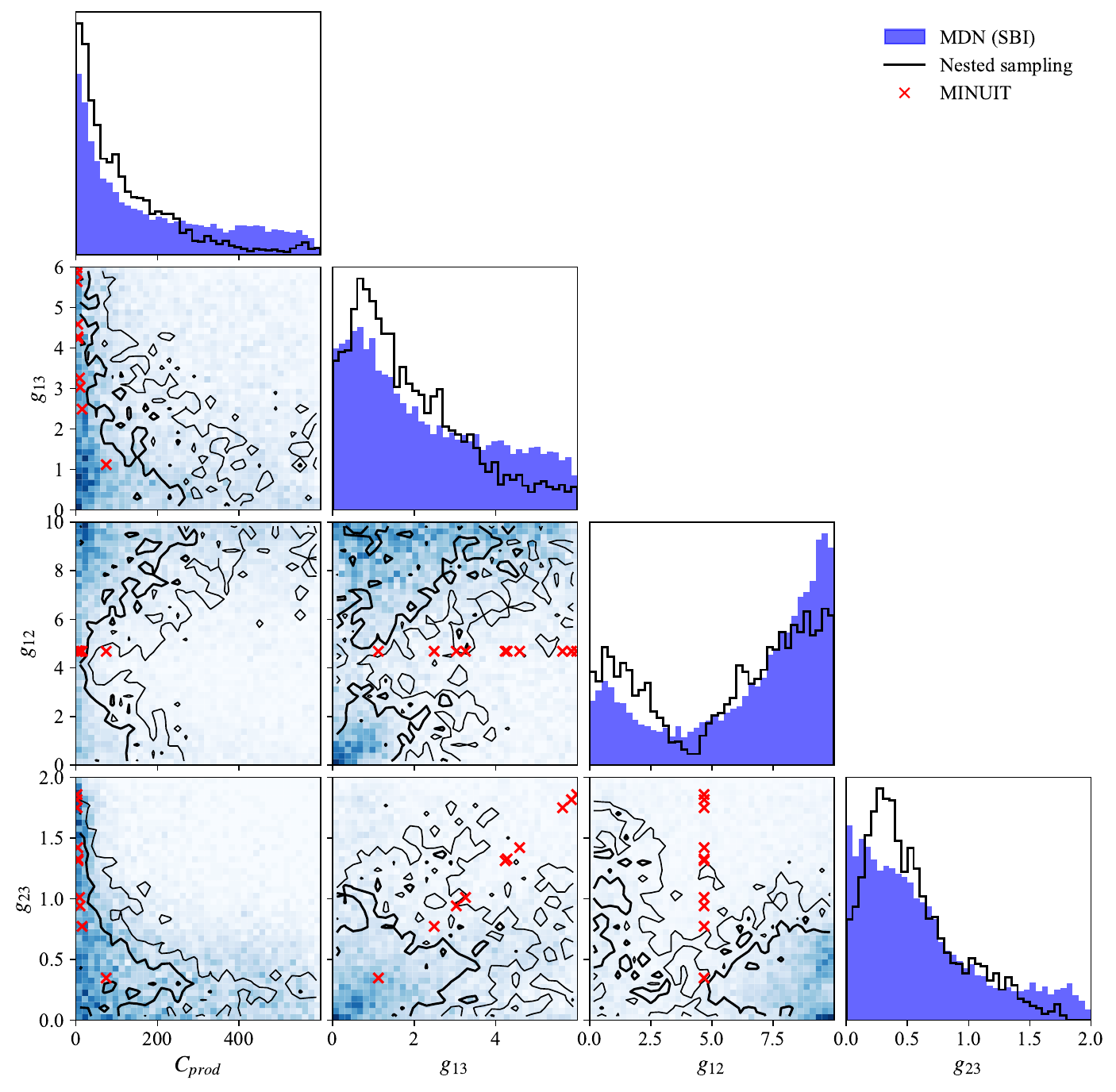}
    \caption{Global Bayesian posterior probability distributions for the effective couplings ($C_{prod}, g_{13}, g_{12}, g_{23}$), generated via Neural Posterior Estimation  conditioned on the BESIII cross-section data ($q_{max} = 0.750$~GeV, $\text{SEED} = 42$)\textcolor{red}{, marginalized over the charmonium
parameters}. Diagonal panels show 1D marginalized posterior distributions;
lower-triangle panels display 2D joint-correlation density maps \textcolor{red}{obtained from the MDN (blue), compared against the nested-sampling results (black). Red crosses mark the degenerate minima found by the MINUIT fit in Sec. \ref{subsec:minuit_fit}, and all axes span the prior intervals of
Eq. (\ref{eq:prior}).}}
    \label{fig:bayesian_corner_plot}
\end{figure}

We begin by sampling $N=\textcolor{red}{50{,}000}$ parameter combinations uniformly from flat priors over the \textcolor{red}{ranges of Sec. \ref{subsec:minuit_fit}}:
\begin{equation}
\textcolor{red}{C_{prod}\in[0,600],\qquad g_{13}\in[0,6],\qquad g_{12}\in[0,10],\qquad g_{23}\in[0,2].}
\label{eq:prior}
\end{equation}
\textcolor{red}{For each parameter set, we also draw the eight charmonium parameters from flat priors,
$A_k\in[0,0.01]$ GeV$^2$, which extends well beyond the fitted amplitudes in  Table \ref{tab:minuit_results}, and
$\phi_k\in(-\pi,\pi]$. Because these quantities are redrawn in every simulation, the MDN directly learns the posterior of
the four couplings marginalized over the charmonium parameters.}
Using our unitarized $3 \times 3$ coupled-channel model with $q_{\mathrm{max}} = 0.750$~GeV, we generate a synthetic dataset of theoretical cross-section lineshapes evaluated by BESIII~\cite{BESIII:2025fph}\textcolor{red}{, adding Gaussian noise with
the uncertainty $\delta_i$ of Eq.~(\ref{eq:chi2}) to each point. In this way, the simulator  draws from the same likelihood function used for the fit in Sec.~\ref{subsec:minuit_fit}}.  Because cross-sections span several orders of magnitude, we \textcolor{red}{divide each input point by its corresponding uncertainty $\delta_i$ and standardize the resulting values to zero
mean and unit variance over the training set}. This ensures the neural network pays equal attention to subtle threshold features as it does to large resonance spikes. The MDN processes the cross-section inputs through  \textcolor{red}{three} hidden layers ($128$ neurons each with ReLU activations) before splitting into three output branches. These branches predict the weights ($\pi_k$), centers ($\boldsymbol{\mu}_k$), and widths ($\boldsymbol{\sigma}_k$) for $K=\textcolor{red}{8}$ Gaussian bell curves\textcolor{red}{, formulated in the variables
$z=\ln[u/(1-u)]$ with $u=\theta/\theta_{max}$, so that every sample remains within the prior
ranges}.
The network is trained by minimizing the Negative Log-Likelihood (NLL) loss \textcolor{red}{on 40{,}000 simulations, reserving 5{,}000 for validation and another 5{,}000 for the
calibration test below. Training stops once the validation loss shows no improvement over 15 epochs.}
To guarantee reproducibility, the random number generators for prior sampling and network initialization were locked ($\text{SEED} = 42$). The \textcolor{red}{validation} loss converged smoothly to $\mathcal{L}_{\text{NLL}} =\textcolor{red}{6.07}$\textcolor{red}{, attained at epoch 120}. Passing the \textcolor{red}{equally scaled} BESIII data vector into the trained network and drawing $10,000$ generative samples yields the Bayesian Corner Plot shown in Figure~\ref{fig:bayesian_corner_plot}\textcolor{red}{,
alongside the nested-sampling posterior discussed below}.
Several key physical insights emerge from this posterior landscape.

The distribution for $g_{13}$ ($D_s \bar{D}_s \to \Omega \bar{\Omega}$) peaks near $g_{13}\approx\textcolor{red}{0.7}$ and \textcolor{red}{exhibits a long tail up to the prior bound;
nested sampling yields $g_{13}=1.50^{+1.81}_{-0.96}$ (median and 68\% interval, see
Table \ref{tab:sbi}).}

\begin{table}[t]
\centering
\caption{\textcolor{red}{Posterior medians and 68\% intervals for the couplings obtained from nested sampling
and the MDN, both marginalized over the charmonium parameters. Shift: difference between the
medians in units of the half-width of the nested-sampling interval. KS: Kolmogorov-Smirnov distance
between the two marginal distributions.}}
\label{tab:sbi}
\textcolor{red}{%
\begin{tabular}{lcccc}
\hline\hline
Parameter & Nested sampling & MDN & Shift & KS \\
\hline
$C_{prod}$ & $83^{+150}_{-65}$         & $142^{+274}_{-118}$      & $+0.55$ & $0.21$ \\
$g_{13}$   & $1.50^{+1.81}_{-0.96}$    & $1.86^{+2.51}_{-1.33}$   & $+0.27$ & $0.13$ \\
$g_{12}$   & $6.67^{+2.40}_{-5.30}$    & $7.59^{+1.81}_{-5.47}$   & $+0.24$ & $0.12$ \\
$g_{23}$   & $0.45^{+0.51}_{-0.25}$    & $0.49^{+0.65}_{-0.35}$   & $+0.11$ & $0.08$ \\
\hline
$C_{prod}\,g_{13}^2$ & $109^{+853}_{-88}$     & $305^{+2437}_{-284}$     & -- & -- \\
$g_{23}/g_{13}$      & $0.29^{+0.61}_{-0.17}$ & $0.24^{+0.84}_{-0.16}$   & -- & -- \\
\hline\hline
\end{tabular}}
\end{table}
 
The distribution for $g_{12}$ (Row 3, Column 3) \textcolor{red}{features two broad modes, one below
$g_{12}\approx3$ and another rising toward the upper prior limit, with a median of $6.67^{+2.40}_{-5.30}$.
Only 6\% of the probability falls within $3<g_{12}<5$, where the MINUIT minimum $g_{12}=4.69$ is found.
Although the chiral quark-soliton prediction, $5.75$--$5.90$ in
our normalization \cite{Yang:2018idi}, falls within the 68\% interval, the BESIII data alone do not
constrain $g_{12}$ within the prior range.} 

$C_{prod}$ \textcolor{red}{peaks near zero and falls off gradually toward the upper prior limit, with 56\%
of the probability lying below $C_{prod} = 100$}. 
There is an inverse relationship between $C_{\mathrm{prod}}$ and $g_{13}$, visible as \textcolor{red}{a continuous curved band that traces the
$C_{prod}\,g_{13}^2\approx$~const relation of Sec. \ref{subsec:minuit_fit} and encompasses the MINUIT minima}. When electromagnetic production is low, the model requires a stronger \textcolor{red}{$D_s\bar
D_s\to\Omega^-\bar\Omega^+$ coupling $g_{13}$}  to match the measured cross section\textcolor{red}{, and vice versa. The same behavior holds for $g_{23}$, which exhibits a rank correlation of $-0.53$ with $C_{prod}$. The posterior estimates for the two
combinations, $C_{prod}\,g_{13}^2=109^{+853}_{-88}$ and $g_{23}/g_{13}=0.29^{+0.61}_{-0.17}$,
are in agreement with Eq.~(\ref{eq:comb}).} This shows that weaker direct photon production should be offset by stronger \textcolor{red}{$D_s\bar D_s$} rescattering to account for the total cross section.

 \textcolor{red}{ With only four couplings of interest,
the posterior can also be evaluated using a standard sampler, providing a direct benchmark to validate the MDN. We
sample the full twelve parameters via nested sampling~\cite{Skilling:2006gxv}, using the dynesty
package~\cite{Speagle:2019ivv} with 500 live points and random-walk slice sampling, adopting the priors
described above and the likelihood $\mathcal L\propto e^{-\chi^2/2}$ with $\chi^2$ given by
Eq.~(\ref{eq:chi2}). Projecting the resulting samples on the couplings $(C_{prod},g_{13},g_{12},g_{23})$ integrates out
the charmonium parameters, yielding the same posterior estimated by the MDN. The run needs
$6.7\times10^5$ likelihood evaluations (15.6~min across 19 parallel processes) and gives an effective sample
size of 3619. The lowest $\chi^2$ reached during sampling is 20.35, above the MINUIT minimum, and the posterior median of
$\chi^2$ is 30.4, roughly 13 units above the minimum, as expected for a twelve-dimensional parameter space.
Table~\ref{tab:sbi} compares the two posteriors. For $g_{13}$, $g_{12}$ and $g_{23}$ the medians
agree within 0.27 of the nested-sampling half-width, and the Kolmogorov-Smirnov (KS) distances are at
most 0.13. The largest difference occurs in $C_{prod}$ (a median shift of 0.55, and a KS distance of 0.21): the MDN assigns slightly more
weight to the tails at large $C_{prod}$ and large $g_{13}$ (Fig.~\ref{fig:bayesian_corner_plot}), rendering the valley somewhat broader; the correlation between $\ln C_{prod}$ and $\ln g_{13}$ is $-0.11$
for the MDN compared to $-0.27$ for nested sampling.}

\textcolor{red}{We also assess the calibration of the MDN using simulation-based
calibration~\cite{Talts:2018zdk} across 1000 simulations of the test set: for each, we draw 199 posterior
samples and record the rank of the true parameters. The true values fall within the central 68\%
interval in 69--71\% of the cases. Moreover, the rank distributions are consistent with uniform for
$C_{prod}$ and $g_{12}$ ($p=0.25$ and 0.70, respectively) and show mild deviations for $g_{13}$ and $g_{23}$
($p=0.012$ and 0.045).}

\textcolor {red} {The simulations and the training take 6.0~min, after
which evaluating the posterior for any data set takes only 1.9~ms, whereas nested sampling must be rerun for each new dataset. The calibration test described above completes in just 2~s with the MDN, compared to an estimated
10.8~days with nested sampling. This represents the advantage of the MDN over a standard sampler: it enables extensive calibration tests,
pseudo-data studies, and rapid updates with future BESIII data at negligible computational cost, with nested sampling serving
the reference posterior for the measured data.}
\subsection{Pole Extraction via Cauchy-Riemann PINN}
\label{subsec:pinn_pole}

Having mapped the parameter space, we proceed to extract the complex $S$-matrix pole position ($E_{pole} = M - i\Gamma/2$) on the unphysical Riemann sheet. 
\textcolor{red}{Once the couplings are fixed, the amplitude
$T=[1-VG]^{-1}V$ is fully determined, and its poles across all channels correspond to the zeros of}
\begin{equation}
\textcolor{red}{D(E)=\det\big[1-V(E)\,G^{(n_\Omega n_\Xi n_{D_s})}(E)\big],}
\label{eq:D}
\end{equation}
\textcolor{red}{where $n_i=0$ ($n_i=1$) selects the physical (unphysical) sheet of channel $i$.
The loop functions are continued to the unphysical sheet by
\begin{equation}
{G_i^{(1)}(s)=G_i(s)+\frac{i\,p_i}{4\pi\sqrt s},
{\rm Im}\,p_i>0,}
\label{eq:GII}    \end{equation}}
\textcolor{red}{which reverses the sign of ${\rm Im}\,G_i$ above the corresponding threshold. We consider four energy
windows between neighbouring thresholds, maintaining a 10~MeV distance from each threshold: below $\Xi\bar\Xi$
(2.450--2.626~GeV), between $\Xi\bar\Xi$ and $\Omega\bar\Omega$ (2.646--3.335~GeV), between
$\Omega\bar\Omega$ and $D_s\bar D_s$ (3.355--3.927~GeV), and above $D_s\bar D_s$
(3.947--4.700~GeV). Within each window we use the sheet reached from the physical axis via continuing
to ${\rm Im}\,E<0$, namely, the sheet with the open channels flipped: $(000)$, $(010)$, $(110)$ and
$(111)$, respectively. These sheets contain the poles that govern the line shape. We locate the zeros
of $D$ directly via Newton iteration started from the local minima of $|D|$ on a grid covering
$\Gamma\le400$~MeV, and verify their number using the argument principle.} 
\textcolor{red}{As an
independent approach requiring only the values of the amplitude along the real axis,} we construct a Cauchy-Riemann Physics-Informed Neural Network (PINN)\textcolor{red}{, and validate it against this
exact pole search}.

\begin{figure}[htbp]
    \centering
    \includegraphics[width=0.82\textwidth]{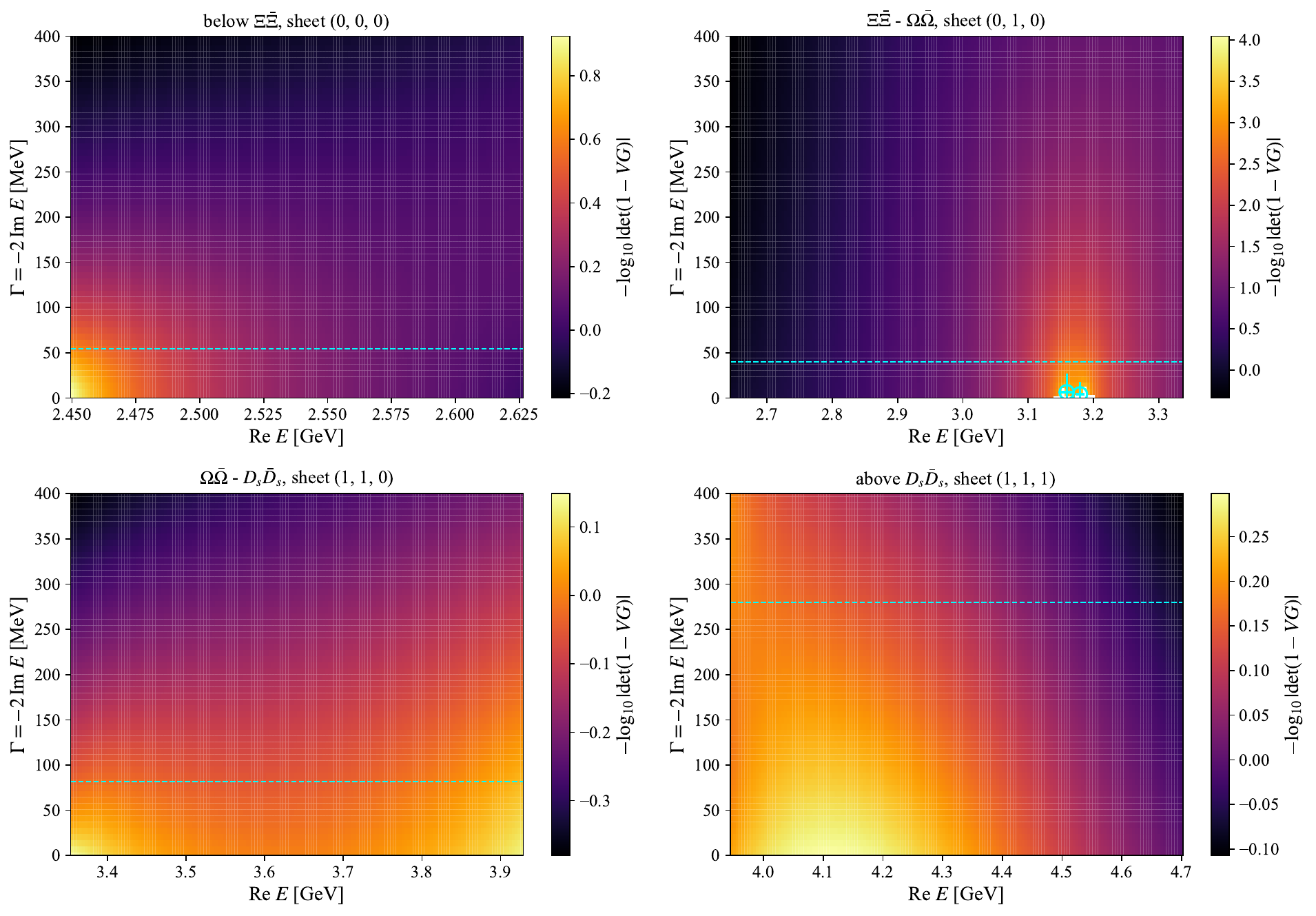}
    \caption{Analytic continuation of \textcolor{red}{$D(E)= \det[1-VG]$, evaluated at the best-fit parameters, into the complex energy plane (Re $E$ vs. $\Gamma$), on the sheet adjacent to the physical axis in each energy window. The color scale represents $-\log_{10}|D|$, such that bright regions correspond to poles of $T$. The dashed line indicates the depth into the complex plane to which the ensemble members agree within 5$\%$. No pole is found near the $D_s\bar{D_s}$ threshold.}}
    \label{fig:pinn_pole_heatmap}
\end{figure}

Standard neural networks struggle to represent singular poles ($T \to \infty$). We solve this by training the network on \textcolor{red}{the determinant $D(E)$ of Eq.~(\ref{eq:D}). Because $D$ has no poles, a pole in any matrix element $T_{ij}$ appears as a smooth zero, $D\to0$, allowing a single function to capture the poles across all three channels.} The PINN \textcolor{red}{is holomorphic by
construction, employing complex weights and biases alongside the entire activation function $\sin$ across three hidden layers of 64 complex neurons. It takes the complex energy {$E$} as
input and outputs the complex value $f\approx D/D_{\rm scale}$, where $D_{\rm
scale}$ is the maximum value of $|D|$ along the real axis within the given energy window}.

The network is optimized \textcolor{red}{with 4000 Adam steps, minimizing only the
data loss
\begin{equation}
\mathcal L_{\rm Data}=\frac{1}{N}\sum_{n=1}^{N}\left|f(E_n)-\frac{D(E_n+i0)}{D_{\rm
scale}}\right|^2,
\label{eq:Ldata}
\end{equation}
evaluated at $N=400$ Chebyshev nodes $E_n$ along the real axis of one window. $f$
is holomorphic by construction; we verify holomorphy via
the residual}
\begin{equation}
\textcolor{red}{\mathcal R_{\rm CR}=\frac{\big\langle(\partial_xu-\partial_yv)^2+(\partial_yu+\partial_xv)^2\big\rangle}
{\big\langle(\partial_xu)^2+(\partial_xv)^2\big\rangle},\qquad E=x+iy,\quad u={\rm Re}\,f,\quad
v={\rm Im}\,f.}
\label{eq:RCR}
\end{equation}

Spatial partial derivatives are computed analytically at every step via automatic differentiation (\texttt{torch.autograd})\textcolor{red}{, at 2000 random points within the network box}. With $\text{SEED} = 42$, the network\textcolor{red}{s} converged to a data loss of \textcolor{red}{$3\times10^{-7}$ to
$3\times10^{-6}$} and a Cauchy-Riemann \textcolor{red}{residual that vanishes to double precision, as
expected for a holomorphic function}.

\textcolor{red}{Training on one window also fixes the corresponding Riemann sheet. Because each window excludes branch
points, the identity theorem ensures that the continuation of $D(E+i0)$ to ${\rm Im}\,E<0$ coincides with $D$ on the
adjacent sheet of that window; we checked that $D$ on this sheet joins the physical axis
continuously. The continuation of real-axis data into the
complex plane remains reliable only to a depth on the order of the window
size. We estimate this uncertainty with an ensemble of eight networks trained with different random seeds
(SEED $=42-49$), and use the networks only where their spread is below 5\% of $D_{\rm scale}$.
Determined solely from the networks, these validity domains reach $\Gamma\le54$, 40, 82 and
280~MeV across the four respective windows. Inside them the error relative to the exact $D$ remains below 15\% of $D_{\rm scale}$;
the ensemble spread underestimates it by up to a factor of 5.5, and the error grows rapidly outside the domains.}

Evaluating \textcolor{red}{$D$ and} the trained \textcolor{red}{networks over the four windows} yields the \textcolor{red}{maps of the adjacent sheets} shown in Figure~\ref{fig:pinn_pole_heatmap}. \textcolor{red}{The exact search identifies zeros exclusively between the $\Xi\bar{\Xi}$ and $\Omega\bar\Omega$
thresholds on sheet $(010)$ (Table~\ref{tab:poles}). The first is a narrow pole} located
\textcolor{red}{near} the real energy axis at:

\begin{equation}
M=\textcolor{red}{3.1868}~\text{GeV},\qquad \Gamma=\textcolor{red}{1.57}~\text{MeV},\qquad
E_B=\textcolor{red}{-158}~\text{MeV},
\label{eq:pole}
\end{equation}
relative to the \textcolor{red}{$\Omega^-\bar\Omega^+$} threshold. 
\textcolor{red}{Our result, $M=3.1868$~GeV and $\Gamma=1.57$~ MeV, is obtained with the first-order expansion of $V_{11}$, see Eq. (\ref{eq:Vthr}). We also examined other truncations. Keeping the zeroth or second order moves the pole within $M=3.180-3.204$~GeV and $\Gamma=1.0-3.5$~MeV, and it remains $\Omega^-\bar\Omega^+$ dominated. 
The second zero,
at $M=3.1540$~GeV, lies on the real axis above the open threshold $\Xi\bar\Xi$, a position forbidden for a physical pole by unitarity. It is a zero of the single-channel factor $1-V_{33}G_3$ and is independent of the inter-channel couplings: it comes from the threshold expansion of $V_{33}$ in Eq.~(\ref{eq:Vthr}), evaluated
783~MeV below the $D_s\bar D_s$ threshold, where the expansion parameter reaches
$2|\vec p_{D_s}|^2/m_\phi^2=2.67$. We therefore regard this solution as an artifact. No zeros are found in the
remaining windows; in particular, no poles appear on sheets $(110)$ or $(111)$ in the vicinity of the $D_s\bar D_s$ threshold, and no complex zeros are present on the physical sheet.}

\begin{table}[t]
\centering
\caption{\textcolor{red}{Zeros of $D(E)$, Eq.~(\ref{eq:D}), on the sheet adjacent to each energy window: exact search and network ensemble. Last column: validity domain of the
networks. $^{a}$Artifact of the threshold expansion of $V_{33}$.}}
\label{tab:poles}
\textcolor{red}{%
\begin{tabular}{lcccccc}
\hline\hline
Window [GeV] & Sheet & \multicolumn{2}{c}{Exact} & \multicolumn{2}{c}{Networks} & Domain\\
 & & $M$ [GeV] & $\Gamma$ [MeV] & $M$ [GeV] & $\Gamma$ [MeV] & $\Gamma$ [MeV]\\
\hline
2.450--2.626 & $(000)$ & -- & -- & -- & -- & $\le54$\\
2.646--3.335 & $(010)$ & $3.1868$ & $1.57$ & $3.179\pm0.007$ & $4\pm13$ & $\le40$\\
             &         & $3.1540^{a}$ & $0$ & $3.159\pm0.010$ & $6\pm20$ & \\
3.355--3.927 & $(110)$ & -- & -- & -- & -- & $\le82$\\
3.947--4.700 & $(111)$ & -- & -- & -- & -- & $\le280$\\
\hline\hline
\end{tabular}}
\end{table}
To quantify the physical coupling of this state to each channel, we evaluate the $S$-matrix pole residues in the complex energy plane \cite{Yu:2018yxl}:
\begin{equation}
g_i^2 = \lim_{\sqrt{s} \to z_R} (\sqrt{s} - z_R) \, T_{ii}(s),
\end{equation}
where $z_R = E_{\mathrm{pole}} = \textcolor{red}{(3.1868-0.0008\,i)}$~GeV. The extracted effective couplings to the three channels are:
\begin{equation}
|g_{D_s}|=\textcolor{red}{0.038~\text{GeV}^{1/2}},\qquad |g_\Xi|=\textcolor{red}{0.265~\text{GeV}^{1/2}},\qquad
|g_\Omega|=\textcolor{red}{8.99~\text{GeV}^{1/2}}.
\label{eq:gres}
\end{equation}

The squared residue for \textcolor{red}{the $\Omega^-\bar\Omega^+$ channel ($|g_\Omega|^2$) exceeds that for $\Xi\bar\Xi$ ($|g_\Xi|^2$), and the $D_s\bar D_s$ residue is smaller}. This compositional hierarchy, combined with the \textcolor{red}{narrow} decay width
($\Gamma=\textcolor{red}{1.57}$~MeV) resulting from \textcolor{red}{its coupling to the open
$\Xi\bar\Xi$ channel}, favors the interpretation of the \textcolor{red}{3.187}~GeV pole as a
dynamically generated, deeply bound \textcolor{red}{$\Omega^-\bar\Omega^+$} state. \textcolor{red}{This state is generated by the attractive $\phi$-exchange potential $V_{11}$ of
Eq.~(\ref{eq:V11}), which contains no free parameters, and its position remains unchanged across all degenerate minima
of Sec.~\ref{subsec:minuit_fit}. Below the $\Omega\bar\Omega$ threshold, $V_{11}$ is replaced by its threshold
expansion, Eq.~(\ref{eq:Vthr}). Because the state lies approximately 586~MeV below the measured energy
range, its position is not constrained by the BESIII data and can be regarded as a prediction of
the model.}
\section{Summary and outlook}
\label{sec: summary}
In this work,  we investigated the underlying reaction dynamics and potential line-shape structures near the $D_s\bar{D}_s$ threshold in the $e^+e^- \to \Omega^-\bar{\Omega}^+$ Born cross section measured by the BESIII collaboration~\cite{BESIII:2025fph}. We constructed a $3 \times 3$ unitarized coupled-channel framework incorporating the $|\Omega^-\bar{\Omega}^+\rangle$, $|\Xi\bar{\Xi}\rangle$, and $|D_s^-\bar{D}_s^+\rangle$ channels. The driving interaction potentials were derived from effective Lagrangians respecting chiral symmetry, heavy quark spin symmetry, and local hidden gauge symmetry, unitarized via the on-shell Bethe-Salpeter equation.
Our main findings are summarized as follows:

By fitting the unitarized scattering amplitude to the experimental cross-section data, \textcolor{red}{ we obtained a good description of the data, $\chi^2/{\rm d.o.f.}=17.66/22$.
However, the fit does not constrain $C_{prod}$, $g_{13}$ and $g_{23}$ individually: they lie along a
flat valley in parameter space where only the combinations $C_{prod}\,g_{13}^2$ and $g_{23}/g_{13}$ are fixed}. The Kaon-exchange coupling $g_{12}=\textcolor{red}{4.69^{+1.54}_{-0.37}}$ agrees with independent predictions from the chiral quark-soliton model (\textcolor{red}{5.75--5.90 in our normalization})~\cite{Yang:2018idi} \textcolor{red}{, but the posterior distribution indicates
that the data alone do not constrain $g_{12}$ within its prior range.}

To overcome the limitations of localized gradient fits, we implemented Simulation-Based Inference (SBI) using a Mixture Density Network (MDN) trained on \textcolor{red}{40{,}000} synthetic
cross-section curves\textcolor{red}{, treating the charmonium parameters as nuisance
parameters. We validated this approach against nested sampling under the same likelihood: the posterior medians
agree within 0.55 of the nested-sampling $1\sigma$ half-width, the MDN intervals exhibit the expected
coverage, and the MDN evaluates the posterior for a new data set in milliseconds rather than minutes}. The predicted Bayesian corner plot mapped out the global parameter landscape, revealing \textcolor{red}{the valley as a continuous band in the $C_{prod}$-$g_{13}$ and $C_{prod}$-$g_{23}$
planes, alongside a broad, bimodal posterior for $g_{12}$}. This  \textcolor{red}{structure arises because
the production amplitude depends on $C_{prod}$, $g_{13}$ and $g_{23}$ effectively only through two
combinations},  explaining why conventional error estimation tools yield large uncertainties.

To extract $S$-matrix poles\textcolor{red}{, we identified the zeros of $\det[1-VG]$ directly on the
Riemann sheets adjacent to the physical axis, and} developed a Cauchy-Riemann Physics-Informed Neural Network (PINN)
\textcolor{red}{ that is holomorphic by construction. Trained only on the
real-axis values of this determinant within a single energy window, which fixes the corresponding Riemann sheet, an
ensemble of networks reproduces the exact zeros to within 6--7.5~MeV and provides the uncertainty of the
continuation. We find no pole close to the $D_s\bar D_s$ threshold. At the best fit, the only physical pole is a narrow
one} at $M=\textcolor{red}{3.1868}$~GeV with $\Gamma=\textcolor{red}{1.57}$~MeV \textcolor{red}{ ($M=3.180-3.204$~GeV for other treatments of $V_{11}$ below threshold)}, corresponding to a
binding energy of $E_B=\textcolor{red}{158}$~MeV below the \textcolor{red}{$\Omega^-\bar\Omega^+$}
threshold.
Evaluating the pole residues yielded a squared coupling \textcolor{red}{$|g_\Omega|^2$ being larger than $|g_\Xi|^2$}, favoring the interpretation of this structure as a dynamically generated, deeply bound 
 \textcolor{red}{$\Omega^-\bar\Omega^+$} state.

Extending the present model to include spin-1 vector meson channels such as $|D_s^{*-}\bar{D}_s^{*+}\rangle$ or heavy baryon pairs like $|\Xi_c\bar{\Xi}_c\rangle$ could provide deeper insights into potential vector-vector and heavy-baryon molecular candidates. 
\textcolor{red}{Measurements of related processes, including $e^+e^-\to\Xi\bar\Xi$ and
$e^+e^-\to D_s^-D_s^+$, would provide independent constraints on $g_{23}$ and $g_{13}$, thereby resolving the degeneracy
found in the present analysis. The posterior of this work can also be used to predict these cross sections.} Comparing these predictions with future measurements will test the universality of our coupled-channel potential. \textcolor{red}{The predicted $\Omega^-\bar\Omega^+$ bound state with $J^{PC}=1^{--}$, may be
investigated through the $e^+e^-\to\Xi\bar\Xi$ near $\sqrt s=3.19$~GeV}. The Cauchy-Riemann PINN methodology
established here offers a \textcolor{red}{framework for analytic continuation when the amplitude is known
only on the real axis}.
It can be applied to other controversial $XYZ$ exotic candidates to extract pole positions on complex Riemann sheets\textcolor{red}{, with direct root-finding as an independent validation
whenever the amplitude is available in closed form}.


\begin{thebibliography}{99}


\bibitem{BESIII:2022kzc}
M.~Ablikim \textit{et al.} [BESIII],
\emph{Study of $e^+e^-\rightarrow \Omega^-\bar{\Omega}^+$ at center-of-mass energies from 3.49 to 3.67~GeV},
\emph{Phys. Rev. D} {\bf 107} (2023)
052003.


\bibitem{BESIII:2026qyk}
M.~Ablikim \textit{et al.} [BESIII],
\emph{Cross sections measurement of $e^+e^-\rightarrow \Xi(1530)^0\bar{\Xi}^0+c.c.$ and search for $\psi(3770)\rightarrow \Xi(1530)^0\bar{\Xi}^0+c.c.$},
\emph{Phys. Rev. D} {\bf 113} (2026) 
072017.


\bibitem{BESIII:2025fph}
M.~Ablikim \textit{et al.} [BESIII],
\emph{Measurement of Born Cross Sections and Effective Form Factors of $e^+e^-\to \Omega^-\bar{\Omega}^+$ from $\sqrt{s}$ = 3.7 to 4.7 GeV}, \emph{JHEP} 02 (2026) 212.


\bibitem{Zhang:2025qmo}
R.~Zhang and X.~Wang,
\emph{Search for charmonium(-like) states decaying into the $\Omega^-\bar{\Omega}^+$ final states},
\emph{Nucl. Phys. B} {\bf 1025} (2026) 117388.


\bibitem{Haidenbauer:2020wyp}
J.~Haidenbauer, U.~G.~Mei{\ss}ner and L.~Y.~Dai,
\emph{Hyperon electromagnetic form factors in the timelike region},
\emph{Phys. Rev. D} {\bf 103} (2021) 
014028.

\bibitem{Dai:2017fwx}
L.~Y.~Dai, J.~Haidenbauer and U.~G.~Mei{\ss}ner,
\emph{Re-examining the $X(4630)$ resonance in the reaction $e^+e^-\rightarrow \Lambda^+_c\bar\Lambda^-_c$},
\emph{Phys. Rev. D} {\bf 96} (2017) 
116001.

\bibitem{Jia:2024ybo}
Z.~S.~Jia, Z.~H.~Zhang, F.~K.~Guo and G.~Li,
\emph{Coupled-channel analysis of the near-threshold e+e-{\textrightarrow}NN{\textasciimacron} cross sections},
\emph{Phys. Rev. D} {\bf 111} (2025) 
054014.

\bibitem{Zhang:2023wmd}
Z.~Zhang and J.~J.~Song,
\emph{Spin density matrix for $\Omega^-$ and its polarization alignment in $\psi(3686) \rightarrow \Omega^-\bar{\Omega}^{+*}$},
\emph{Chin. Phys. C} {\bf 47} (2023) 
093101.

\bibitem{Salnikov:2023qnn}
S.~G.~Salnikov and A.~I.~Milstein,
\emph{Near-threshold resonance in $e^+e^-\rightarrow\Lambda_c\bar{\Lambda}_c$ process},
\emph{Phys. Rev. D} {\bf 108} (2023) 
L071505.

\bibitem{Milstein:2022bfg}
A.~I.~Milstein and S.~G.~Salnikov,
\emph{Final-state interaction in the process $e^+e^-\rightarrow \Lambda_c\bar{\Lambda}_c$},
\emph{Phys. Rev. D } {\bf 105} (2022) 
074002.

\bibitem{Hyodo:2011ur}
T.~Hyodo and D.~Jido,
\emph{The nature of the Lambda(1405) resonance in chiral dynamics},
\emph{Prog. Part. Nucl. Phys.} {\bf 67} (2012) 55-98.


\bibitem{Doring:2025sgb}
M.~D{\"o}ring, J.~Haidenbauer, M.~Mai and T.~Sato,
\emph{Dynamical coupled-channel models for hadron dynamics},
\emph{Prog. Part. Nucl. Phys.} {\bf 146} (2026) 104213.


\bibitem{Shen:2022zvd}
C.~W.~Shen, Y.~h.~Lin and U.~G.~Mei{\ss}ner,
\emph{$P_{cc}^N$ states in a unitarized coupled-channel approach}, \emph{Eur. Phys. J. C}
 {\bf 83} (2023) 
 70.


\bibitem{Qi:2023gwb}
J.~J.~Qi, Z.~Y.~Wang, Z.~F.~Zhang and X.~H.~Guo,
\emph{The properties of the $S$-wave $D_s\bar{D}_s$ bound state},
\emph{Chin. Phys. C} {\bf 50} (2026) 
022001.




\bibitem{Kim:2025ado}
H.~J.~Kim and H.~C.~Kim,
\emph{Production mechanism of doubly charmed exotic mesons $T_{cc}$},
\emph{Phys. Rev. D} {\bf 112} (2025) 
094025.



\bibitem{Qi:2021iyv}
J.~J.~Qi, Z.~Y.~Wang, Z.~F.~Zhang and X.~H.~Guo,
\emph{Studying the ${\bar{D}}_1K$ molecule in the Bethe{\textendash}Salpeter equation approach},
\emph{Eur. Phys. J. C} {\bf 81} (2021) 
639.



\bibitem{Zhang:2024fxy}
Z.~H.~Zhang, T.~Ji, X.~K.~Dong, F.~K.~Guo, C.~Hanhart, U.~G.~Mei{\ss}ner and A.~Rusetsky,
\emph{Predicting isovector charmonium-like states from X(3872) properties},
\emph{JHEP} {\bf 08} (2024) 130.



\bibitem{Moir:2016srx}
G.~Moir, M.~Peardon, S.~M.~Ryan, C.~E.~Thomas and D.~J.~Wilson,
\emph{Coupled-Channel $D\pi$, $D\eta$ and $D_{s}\bar{K}$ Scattering from Lattice QCD},
\emph{JHEP} {\bf 10} (2016) 011.

\bibitem{Lih:2026xgi}
C.~C.~Lih and C.~Q.~Geng,
\emph{Time-like electromagnetic form factors of {\ensuremath{\Lambda}},~{\ensuremath{\Sigma}}, and {\ensuremath{\Xi}}+ in a light-front quark model},
\emph{Phys. Rev. C} {\bf 113} (2026) 
055501.

\bibitem{Wang:2025dur}
\textcolor{red}{
X.~Wang, X.~Liu and Y.~Gao,
\emph{Colloquium: Hadron production in open-charm meson pairs at $e^+e^-$ colliders},
\emph{Rev. Mod. Phys.} \textbf{98}, 
(2026) 021001}

\bibitem{Pilloni:2016obd}
A.~Pilloni \textit{et al.} [JPAC],
\emph{Amplitude analysis and the nature of the Z$_c$(3900)},
\emph{Phys. Lett. B} {\bf 772} (2017) 200-209.



\bibitem{Sombillo:2021rxv}
D.~L.~B.~Sombillo, Y.~Ikeda, T.~Sato and A.~Hosaka,
\emph{Model independent analysis of coupled-channel scattering: A deep learning approach},
\emph{Phys. Rev. D} {\bf 104} (2021) 
036001.

\bibitem{Ng:2021ibr}
L.~Ng \textit{et al.} [Joint Physics Analysis Center and JPAC],
\emph{Deep learning exotic hadrons},
\emph{Phys. Rev. D} {\bf 105} (2022) 
L091501.

\bibitem{Aarts:2025gyp}
G.~Aarts, K.~Fukushima, T.~Hatsuda, A.~Ipp, S.~Shi, L.~Wang and K.~Zhou,
\emph{Physics-driven learning for inverse problems in quantum chromodynamics},
\emph{Nature Rev. Phys.} {\bf 7} (2025) 
154-163.

\bibitem{Liu:2022uex}
J.~Liu, Z.~Zhang, J.~Hu and Q.~Wang,
\emph{Study of exotic hadrons with machine learning},
\emph{Phys. Rev. D} {\bf 105} (2022) 
076013.


\bibitem{Frohnert:2025usi}
F.~Frohnert, D.~L.~B.~Sombillo, E.~van Nieuwenburg and P.~Emonts,
\emph{Learning pole structures of hadronic states using predictive uncertainty estimation},
\emph{Phys. Rev. D} {\bf 113} (2026) 
056019.

\bibitem{Cranmer:2019eaq}
K.~Cranmer, J.~Brehmer and G.~Louppe,
\emph{The frontier of simulation-based inference},
\emph{Proc. Nat. Acad. Sci.} {\bf 117} (2020) 
30055-30062.

\bibitem{ParticleDataGroup:2026mpi}
\textcolor{red}{
F.~Takahashi \textit{et al.} [Particle Data Group],
\emph{Review of Particle Physics},
\emph{Int. J. Mod. Phys. A} \textbf{41} 
(2026) 2630011.}

\bibitem{Oset:2005ag}
E.~Oset, D.~Cabrera, V.~K.~Magas, L.~Roca, S.~Sarkar, M.~J.~Vicente Vacas and A.~Ramos,
\emph{Chiral dynamics of baryon resonances and hadrons in a nuclear medium},
\emph{Pramana J. Phys.} {\bf 66} (2006) 731-752.

\bibitem{Mei:2022msh}
\textcolor{red}{X.~H.~Mei, Z.~Yu, M.~Song, J.~Y.~Guo, G.~Li and X.~Luo,
\emph{Explanation of Y(4630) as a hadronic resonant state},
\emph{Chin. Phys. C} \textbf{47} 
(2023) 033104.}

\bibitem{Yu:2018yxl}
Q.~X.~Yu, R.~Pavao, V.~R.~Debastiani and E.~Oset,
\emph{Description of the $\Xi _c$ and $\Xi _b$ states as molecular states},
\emph{Eur. Phys. J. C} {\bf 79} (2019) 
167.

\bibitem{Korpa:2011qg}
C.~L.~Korpa,
\emph{Effects of consistent and inconsistent isobar coupling in the nuclear medium},
\emph{Phys. Rev. C} {\bf 85} (2012) 014601.



\bibitem{Oset:2010tof}
E.~Oset and A.~Ramos,
\emph{Dynamically generated resonances from the vector octet-baryon octet interaction},
\emph{Eur. Phys. J. A} {\bf 44} (2010) 445-454.


\bibitem{Xiao:2013yca}
C.~W.~Xiao, J.~Nieves and E.~Oset,
\emph{Combining heavy quark spin and local hidden gauge symmetries in the dynamical generation of hidden charm baryons},
\emph{Phys. Rev. D} {\bf 88} (2013) 056012.

\bibitem{Lyu:2026rsm}
W.~T.~Lyu, L.~R.~Dai and E.~Oset,
\emph{Role of $\Xi (1690)$ in the $J/\psi \rightarrow \Xi ^0{\bar{\Lambda }}K^0$ reaction},
\emph{Eur. Phys. J. C} {\bf 86} (2026) 
726.

\bibitem{Yan:1992gz}
T.~M.~Yan, H.~Y.~Cheng, C.~Y.~Cheung, G.~L.~Lin, Y.~C.~Lin and H.~L.~Yu,
\emph{Heavy quark symmetry and chiral dynamics},
\emph{Phys. Rev. D} {\bf 46} (1992) 1148-1164.
\emph{Erratum: Phys. Rev. D} {\bf 55} (1997) 5851.

\bibitem{Bayar:2022dqa}
\textcolor{red}{
M.~Bayar, A.~Feijoo and E.~Oset,
\emph{X(3960) seen in $D_s^+D_s^-$ as the X(3930) state seen in $D^+D^-$},
\emph{Phys. Rev. D} \textbf{107} 
(2023) 034007.}

\bibitem{Gulmez:2016scm}
\textcolor{red}{
D.~G{\"u}lmez, U.~G.~Mei{\ss}ner and J.~A.~Oller,
\emph{A chiral covariant approach to $\rho\rho$ scattering},
\emph{Eur. Phys. J. C} \textbf{77} 
 (2017) 460.}

\bibitem{Epelbaum:2008ga}
E.~Epelbaum, H.~W.~Hammer and U.~G.~Meissner,
\emph{Modern Theory of Nuclear Forces},
\emph{Rev. Mod. Phys.} {\bf 81} (2009) 1773-1825.


\bibitem{Gamermann:2009uq}
D.~Gamermann, J.~Nieves, E.~Oset and E.~Ruiz Arriola,
\emph{Couplings in coupled channels versus wave functions: application to the X(3872) resonance},
\emph{Phys. Rev. D} {\bf 81} (2010) 014029.


\bibitem{Montana:2022inz}
G.~Monta{\~n}a, A.~Ramos, L.~Tolos and J.~M.~Torres-Rincon,
\emph{X(3872), X(4014), and their bottom partners at finite temperature},
\emph{Phys. Rev. D} {\bf 107} (2023) 
054014.

\bibitem{Feijoo:2022rxf}
A.~Feijoo, W.~F.~Wang, C.~W.~Xiao, J.~J.~Wu, E.~Oset, J.~Nieves and B.~S.~Zou,
\emph{A new look at the $P_{cs}$ states from a molecular perspective},
\emph{Phys. Lett. B} {\bf 839} (2023) 137760.

\bibitem{Feijoo:2024bvn}
A.~Feijoo, M.~Korwieser and L.~Fabbietti,
\emph{Relevance of the coupled channels in the $\phi$p and ${\rho}^0$p correlation functions},
\emph{Phys. Rev. D} {\bf 111} (2025) 
014009.

\bibitem{Li:2026umb}
M.~Y.~Li, G.~Y.~Wang, N.~C.~Wei, D.~M.~Li and E.~Wang,
\emph{Probing the isospin structure and low-lying resonances in \(\Lambda_{c}^{+}\rightarrow n{\overline{K}}^{0}\pi^{+}\) decays},
\emph{Phys. Rev. D} {\bf 113} (2026) 
094013.


\bibitem{Wang:2025jcq}
L.~L.~Wang, X.~M.~Zhao, X.~H.~Liu and M.~J.~Yan,
\emph{New spectrum of charm-strange meson with constituent quark model \(c\overline{s}\) contributions},
\emph{Phys. Lett. B} {\bf 873} (2026) 140164.

\bibitem{Rahmani:2026cjb}
S.~Rahmani,
\emph{Dynamical generation of strange molecular states via vector
meson exchange}
\emph{Eur. Phys. J. C} 
{\bf 86} (2026) 
884.



\bibitem{Wang:2024fsz}
Z.~Y.~Wang, Y.~S.~Li and S.~Q.~Luo,
\emph{Scalar resonance contributions in the $D_{s1}(2460)^+ \rightarrow D_s^+\pi^+\pi^-$ reaction},
\emph{Phys. Rev. D} {\bf 111} (2025) 
076009.

\bibitem{Nishibuchi:2023acl}
T.~Nishibuchi and T.~Hyodo,
\emph{Analysis of the {\ensuremath{\Xi}}(1620) resonance and K{\textasciimacron}{\ensuremath{\Lambda}} scattering length with a chiral unitary approach},
\emph{Phys. Rev. C} {\bf 109} (2024) 
015203.


\bibitem{Rahmani:2026gcx}
S.~Rahmani,
\emph{Analysis of the $e^+e^- \rightarrow \eta J/\psi $ cross section within a unitarized coupled-channel approach},
\emph{Eur. Phys. J. C} {\bf 86} (2026) 
513.


\bibitem{Rahmani:2025uut}
S.~Rahmani, W.~Liang, Y.~W.~Peng, Y.~Lu, D.~L.~Yao and C.~W.~Xiao,
\emph{Role of $a_0(980)$ in the decays $D^{0} \rightarrow K^{+} K^{-} \eta$ and $\pi^{+} \pi^{-} \eta$},
\emph{Phys. Rev. D} {\bf 112} (2025) 
036001.

\bibitem{Oller:2000fj}
\textcolor{red}{J.~A.~Oller and U.~G.~Meissner,
\emph{Chiral dynamics in the presence of bound states: kaon nucleon interactions revisited},
\emph{Phys. Lett. B} \textbf{500},  (2001) 263-272.}

\bibitem{Guo:2005wp}
\textcolor{red}{
F.~K.~Guo, R.~G.~Ping, P.~N.~Shen, H.~C.~Chiang and B.~S.~Zou,
\emph{$S$ wave $K\pi$ scattering and effects of $\kappa$ in $J/\psi \to \bar{K}^{*0} (892) K^+ \pi^-$},
\emph{Nucl. Phys. A} \textbf{773} (2006) 78-94.}











\bibitem{Li:2023olv}
H.~P.~Li, G.~J.~Zhang, W.~H.~Liang and E.~Oset,
\emph{Theoretical interpretation of the $\Xi (1620)$ and $\Xi (1690)$ resonances seen in $\Xi _c^+ \rightarrow \Xi ^- \pi ^+ \pi ^+$ decay},
\emph{Eur. Phys. J. C} {\bf83} (2023) 
954.


\bibitem{Xiao:2024kfq}
C.~W.~Xiao and J.~J.~Wu,
\emph{Searching for bound states in the open strangeness systems},
\emph{Eur. Phys. J. A} {\bf 61} (2025) 
179


\bibitem{Yang:2018idi}
G.~S.~Yang and H.~C.~Kim,
\emph{Meson{\textendash}baryon coupling constants of the SU(3) baryons with flavor SU(3) symmetry breaking},
\emph{Phys. Lett. B} {\bf 785} (2018) 434-440.

\bibitem{Burton:2021tsd}
C.~Burton, S.~Stubbs and P.~Onyisi,
\emph{Mixture density network estimation of continuous variable maximum likelihood using discrete training samples},
\emph{Eur. Phys. J. C} {\bf 81} (2021) 
662.

\bibitem{Albert:2024zsh}
J.~Albert \textit{et al.} [DarkMachines High Dimensional Sampling Group],
\emph{A comparison of Bayesian sampling algorithms for high-dimensional particle physics and cosmology applications},
\emph{Comput. Phys. Commun.} {\bf 315} (2025) 109756.


\bibitem{Skilling:2006gxv}
\textcolor{red}{
J.~Skilling,
\emph{Nested sampling for general Bayesian computation},
\emph{Bayesian Analysis} \textbf{1} 
(2006) 833-859.}

\bibitem{Speagle:2019ivv}
\textcolor{red}{
J.~S.~Speagle,
\emph{dynesty: a dynamic nested sampling package for estimating Bayesian posteriors and evidences},
\emph{Mon. Not. Roy. Astron. Soc.} \textbf{493} 
(2020) 3132-3158.}


\bibitem{Talts:2018zdk}
\textcolor{red}{S.~Talts, M.~Betancourt, D.~Simpson, A.~Vehtari and A.~Gelman,
\emph{Validating Bayesian Inference Algorithms with Simulation-Based Calibration},
arXiv:1804.06788 [stat.ME].}













\end{thebibliography}
\end{document}